\documentclass[12pt]{iopart}

\expandafter\let\csname equation*\endcsname\relax

\expandafter\let\csname endequation*\endcsname\relax

\usepackage{iopams}  
\usepackage{graphicx} 
\usepackage{subcaption} 
\usepackage{xcolor}
\usepackage{hyperref}
\usepackage{booktabs}

\usepackage{physics}

\DeclareOldFontCommand{\rm}{\normalfont\rmfamily}{\mathrm}

\begin{document}

\title{Amorphous Boron Nitride as an Ultrathin Copper Diffusion Barrier for Advanced Interconnects}

\author{Onurcan Kaya$^{1,2,3}$, Hyeongjoon Kim$^{4, 5}$, Byeongkyu Kim$^{6}$, Thomas Galvani$^{1}$, Luigi Colombo$^{7}$, Mario Lanza$^{8}$, Hyeon-Jin Shin$^{9}$, Ivan Cole$^{10}$, Hyeon Suk Shin$^{5,11,12}$, and Stephan Roche$^{1,13}$ }

\address{$^1$ Catalan Institute of Nanoscience and Nanotechnology (ICN2), CSIC and BIST, Campus UAB, Bellaterra, 08193, Barcelona, Spain}
\address{$^2$ School of Engineering, RMIT University, Melbourne, Victoria, 3001, Australia }
\address{$^3$ Department of Electronic Engineering, Universitat Autonoma de Barcelona (UAB), Campus UAB, Bellaterra, 08193 Barcelona, Spain}
\address{$^4$ Department of Chemistry, Ulsan National Institute of Science and Technology,  Ulsan 44919, Republic of Korea} 
\address{$^5$ Center for 2D Quantum Heterostructures, Institute for Basic Science (IBS), Suwon16419, Republic of Korea}
\address{$^6$School of Semiconductor Materials and Devices Engineering, Ulsan National Institute of Science and Technology (UNIST), Ulsan 44919, Republic of Korea }
\address{$^7$ CNMC, LLC, Dallas, TX, 75248, USA}
\address{$^8$ Department of Materials Science and Engineering, National University of Singapore, 117575, Singapore}
\address{$^9$ Department of Semiconductor Engineering, School of Electrical Engineering and Computer Science, Gwangju Institute of Science and Technology (GIST), Republic of Korea}
\address{$^{10}$ School of Engineering, The Australian National University , Canberra, ACT ,2600} 
\address{$^{11}$ Department of Energy Science, Sungkyunkwan University (SKKU), Suwon 16419, Republic of Korea}
\address{$^{12}$ Department of Chemistry, Sungkyunkwan University (SKKU), Suwon 16419, Republic of Korea}
\address{$^{13}$ ICREA Institucio Catalana de Recerca i Estudis Avancats, 08010 Barcelona, Spain}

\ead{stephan.roche@icn2.cat}

\begin{abstract}
This study focuses on amorphous boron nitride ($\rm \alpha$-BN) as a novel diffusion barrier for advanced semiconductor technology, particularly addressing the critical challenge of copper diffusion in back-end-of-line (BEOL) interconnects. Owing to its ultralow dielectric constant and robust barrier properties, $\rm \alpha$-BN is examined as an alternative to conventional low-k dielectrics. The investigation primarily employs theoretical modelling, using a Gaussian Approximation Potential, to simulate and understand the atomic-level interactions. This machine learning-based approach allows the performance of realistic simulations of amorphous structure of $\rm \alpha$-BN, enabling the exploration of the impact of different film morphologies on barrier efficacy. Furthermore, we studied the electronic and optical properties of the films using a simple Tight-Binding model. In addition to the theoretical studies, we performed diffusion studies of copper through PECVD $\rm \alpha$-BN on Si. The results from both the theoretical and experimental investigations highlight the potential of $\rm \alpha$-BN as a highly effective diffusion barrier, suitable for integration in nanoelectronics. This research shows that $\rm \alpha$-BN is a promising candidate for BEOL interconnects combining the machine learned MD simulations with PECVD measurements.

\end{abstract}

\section{Introduction}
The continuous down-scaling of devices and architectures has been the major goal of the electronic industry for decades in order to increase performance, chip density, and reduce power consumption \cite{SHAMIRYAN200434, Hong2020, Liu2004plasma}. Interconnect technology is also experiencing significant challenges as devices are scaled down, with sustained efforts to decrease the resistance of metals and to lower the dielectric constant of intermetal dielectrics (IMDs), for optimised control of signal delays such as RC delay. The diffusion barrier that prevents the interaction between metals and IMD is also becoming thinner and is required to have excellent blocking properties. Ultimately, the technology aims to minimise barrier thickness in BEOL. Two-dimensional (2D) materials, such as $\rm MoS_2$ and $\rm h-BN$ have recently garnered attention as potential barrier materials. Indeed, recent studies have shown that $\rm MoS_2$, when synthesized through Atomic Layer Deposition (ALD), can act as an effective Cu diffusion barrier, potentially reducing barrier thickness down to 1-2 nm\cite{Lo2018, Deijkers2023,zhao2019incorp}. Furthermore, the integration of dopants into $\rm MoS_2$ such as Niobium (Nb) has been found to enhance its ability to block Cu atom diffusion, indicating its suitability for sub-5 nm technology nodes. However, while these 2D materials could act as efficient diffusion barriers, they also have a higher dielectric constant which would add to the RC delay. \\
Hexagonal boron nitride ($\rm h$-BN), on the other hand, could be a better choice due to its low dielectric constant and higher bandgap. However, the practical application of these 2D materials is confronted with significant challenges. Integrating these materials into existing semiconductor manufacturing processes, especially developing a BEOL-compatible growth process for deposition on dielectrics remains a major challenge \cite{Maestre_2021, Ranjan2023molecular, Jiang2017dielectric}. Moreover, the presence of defects in 2D materials, such as vacancy defects in hBN and $\rm MoS_2$, can lead to Cu accumulation, creating diffusion pathways that compromise the barrier's effectiveness \cite{Ahmed_2022}. Amidst the ongoing quest for materials that can meet the stringent demands of advanced semiconductor technologies, amorphous boron nitride ($\rm \alpha$-BN) emerges as a viable candidate. In contrast to h-BN, $\rm \alpha$-BN does not suffer from dangling bonds and grain boundaries. In addition to its low temperature growth, $\rm \alpha$-BN films can be uniformly synthesized on various substrates (such as metals and dielectrics) with a controllable thickness. This makes $\rm \alpha$-BN a desirable dielectric material and diffusion barrier \cite{Glavin2014, Glavin2016}. Besides, $\rm \alpha$-BN has been reported to have a low dielectric constant as a result of poor dipole alignment due to its amorphous nature \cite{SattariEsfahlan2023}. While Hong et al.\cite{Hong2020} reported an ultralow dielectric constant of 1.7 for 3 nm PECVD-grown $\rm \alpha$-BN. Glavin et al.\cite{Glavin2016} pointed out that homonuclear bonds may open conductive pathways and increase the dielectric constant. They measured a device dielectric constant of atomic layer deposition (ALD)-grown $\rm \alpha$-BN of 5.9 with a band gap of 4.5 eV \cite{Glavin2016}. Similarly, the dielectric constant of ALD grown $\rm \alpha$-BN at 65 - 250 ºC is between 4.3 - 9.0 which increases with faster cooling. They observed that faster cooling leads to N deficiency and higher oxygen contamination, causing midgap states and higher dielectric constant \cite{chen2023tailoring}. Lin et al.\cite{Lin2022} achieved a dielectric constant of 2.1 for sub-10 nm $\rm \alpha$-BN films by optimising the growth process and using borazine as a precursor. Meanwhile, Abbas et al.\cite{Abbas2018} reported dielectric constants of 2.61 - 5.88 for sputtered $\rm \alpha$-BN films with thicknesses ranging between 12 nm and 800 nm\cite{Abbas2018}. Due to their low dielectric constant, good mechanical properties, and thermal stability, $\rm \alpha$-BN films have already been shown to act as efficient metal diffusion barriers. While Hong et al. \cite{Hong2020} showed that a 3 nm $\rm \alpha$-BN film can stop Co diffusion at 600 ºC, Kim et al. \cite{Kim2023} found that 7-nm-$\rm \alpha$-BN film can stop Cu diffusion up to 500 ℃. Hence, $\rm \alpha$-BN seems to be the ideal dielectric material due to its ultralow dielectric constant and great barrier properties. However, as shown in the literature \cite{Glavin2016, Lin2022, chen2023tailoring, Kaya2023, kaya2023impact}, the properties and performance of $\rm \alpha$-BN are determined by their morphology. Hence, ALD $\rm \alpha$-BN with the correct composition could be an ideal dielectric material due to the potential low dielectric constant and excellent metal barrier properties.\\
An exhaustive experimental investigation of the morphology and properties of $\rm \alpha$-BN  would be quite time consuming and expensive. On the other hand, atomistic simulations would allow us to study materials at the atomic level and can provide faster insights into the relation between morphology and key parameters of materials with moderate computational cost. Notwithstanding, implementing an accurate modelling of highly disordered materials such as $\rm \alpha$-BN is quite a challenging task. Due to the lack of long-range order, the simulation box of amorphous materials should be quite large. This becomes quickly prohibitive for {\it ab initio} calculations. On the other hand, simulations using classical Molecular Dynamics (MD) with empirical potentials can deal with such large samples. However, due to simplifications and assumptions used to develop these interatomic potentials, classical MD cannot capture the diverse atomic environments of the amorphous material. These challenges can be overcome using machine learning-based approaches. Classical MD simulations with machine learning-based interatomic potential trained over {\it ab initio} calculations can deal with very large simulation boxes with {\it ab initio} accuracy \cite{Bartok2010, Bartok2013, Villena2024, Deringer2018}. \\
Here, the relationship between the morphology of $\rm \alpha$-BN structures and their barrier properties is assessed both theoretically and experimentally. First, we develop a Gaussian Approximation Potential (GAP) model to describe the interactions between $\rm \alpha$-BN and Cu atoms. Later, we generate $\rm \alpha$-BN films with different thicknesses, and by employing different parameters, we investigate their resulting morphologies and barrier properties. We also study the mechanical properties of generated $\rm \alpha$-BN samples, as well as their electronic and dielectric properties. Finally, we present an experimental investigation of the barrier performance of PECVD-grown 3-, 5-, and 7-nm thick $\rm \alpha$-BN films against Cu diffusion to the Si-based substrate and compare it with the simulations. 
\section{Methods}
This study combines both computational models and experimental techniques to evaluate the diffusion barrier potential of ultrathin amorphous boron nitride ($\rm \alpha$-BN) films against copper (Cu). In our computational investigation, we used molecular dynamics (MD) simulations and a tight binding (TB) model to generate $\rm \alpha$-BN samples with different qualities, calculate the mechanical and optical properties, and evaluate their diffusion barrier properties. Experimental efforts focused on the barrier properties of PECVD-grown $\rm \alpha$-BN films of different thicknesses by detecting the presence of copper silicide crystals. Both experimental and computational work provide a good understanding of the effectiveness of $\rm \alpha$-BN as a Cu diffusion barrier in future electronic applications.
\subsection{Molecular Dynamics Simulations}
MD simulations use Newton's law of motion to follow the trajectories of atoms through time. The accuracy and reliability of MD simulations largely depends on the accuracy of interatomic potentials. While there are several empirical potentials for $\rm \alpha$-BN and Cu separately, there is no potential that takes both BN and Cu into account. Moreover, empirical potentials provide insufficient accuracy for such complex systems \cite{Bartok2010, Bartok2013, Deringer2017}. Machine learning based interatomic potentials that utilise a large dataset of quantum-mechanically computed reference data for structures, energies and forces are quite useful for disordered systems since they enable us to study large numbers of atoms with an accuracy comparable to DFT and a reasonable computing cost. Therefore, we trained a Gaussian Approximation Potential for BNCu systems that is based on DFT-generated energy and force data of a large database. The dataset contains isolated atoms, molecular compounds, crystalline structures, heterostructures and random structures. To ensure accuracy beyond nearest neighbours, the GAP employs SOAP descriptors with a cutoff r$_{\rm cut}$=7 $\rm \AA$, which spans first, second, and higher shells in $\rm \alpha$-BN and include intermediate medium-range interactions in $\rm \alpha$-BN \cite{Elliott1991}. Details of training and validation of the GAP model are presented in \ref{a:GAP}.\\
\begin{figure}[htb!]
    \centering
    \includegraphics[width=\textwidth]{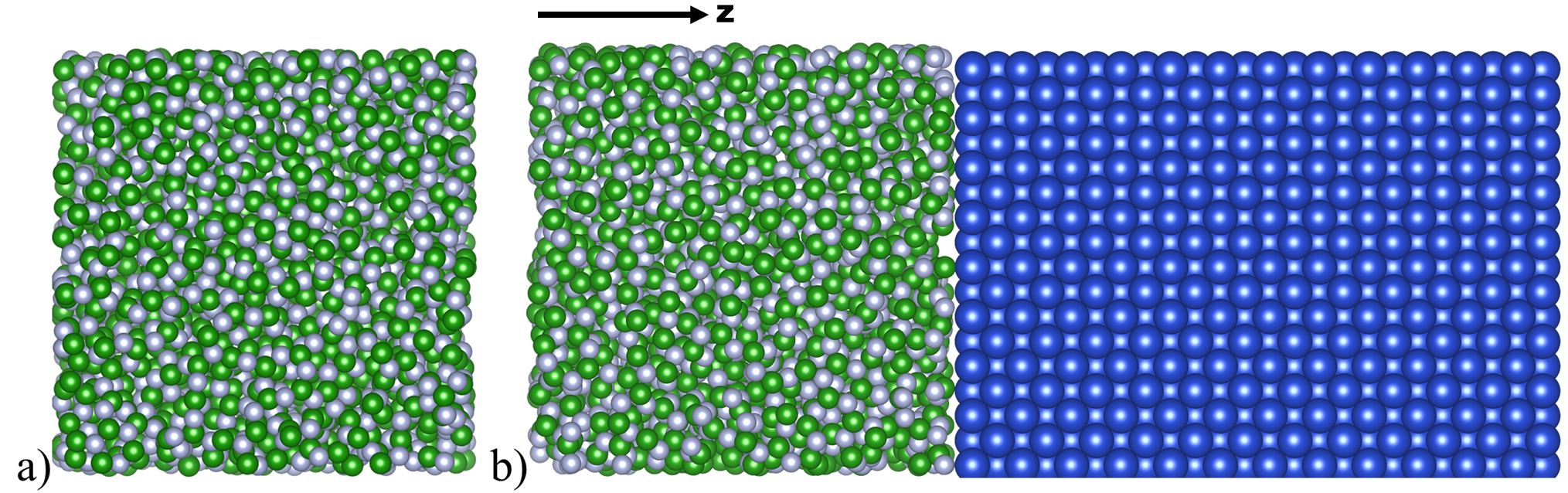}
    \caption{a) A sample of simulation box of 3-nm thick $\rm \alpha$-BN films generated by a melt-quench protocol of the type used to calculate mechanical and dielectric properties; this structure was used for calculating mechanical and dielectric properties, b) a sample of simulation cells of 3-nm thick $\rm \alpha$-BN/Cu heterostructures used for barrier calculations, where B, N, and Cu atoms are represented as green, white, and blue, respectively. Note that The z-axis, corresponding to the film thickness, is displayed horizontally in the figure for clarity. 2.5 nm thick vacuum spaces are applied to both of end of simulation cells, which are not shown here.}
    \label{fig:sample}
\end{figure}
To generate $\rm \alpha$-BN films with different qualities, we employed a melt-quench protocol. This protocol is an efficient way to generate amorphous structures and glasses in MD and has been used extensively by community \cite{Kaya2023, kaya2023impact,Deringer2017}. In this protocol, an equal number of B and N atoms were placed in a cube simulation box which is periodic in all directions. However, we placed a 25 $\rm \AA$-thick vacuum region at both the top and bottom of the film to prevent boundary interactions and mimic thin films.  After a minimisation and an equilibration run at 5000 K for 20 ps with a timestep of 0.25 fs (same timestep used in all simulation), all films were cooled down with a constant cooling rate (CR) of 5 - 100 K/ps under NVT ensemble (constant volume-constant temperature) to systematically study the effect of cooling rates on film morphology, capturing variations in the degree of disorder characteristic of amorphous materials. Later all films are relaxed under NVE (constant volume-constant energy) and equilibrated under NVT ensemble for 20 ps each. Here, the initial density of each cube (before the addition of the vacuum regions) was fixed at approximately 2.1 $\rm g/cm^3$, the experimentally reported value for $\rm \alpha$-BN films \cite{Hong2020}. The number of atoms in 3-nm, 5-nm and 7-nm thick films are respectively 3000, 13000 and 37000 to maintain the selected density. Upon generating the films, several morphological features are investigated to evaluate the quality of samples. We checked the presence of $\rm sp^2$ and $\rm sp^3$ atoms, homonuclear bonds, nanovoids, as well as the density of the samples. We generated 5 samples at each thickness with each cooling rate to get statistically meaningful data. A 3-nm thick $\rm \alpha$-BN sample generated using this protocol is shown in Fig.\ref{fig:sample}-a.  \\
Later, we investigated the mechanical properties of $\rm \alpha$-BN thin films. The elastic constants of $\rm \alpha$-BN thin films were calculated by computing the stress fluctuations and the Born matrix of $\rm \alpha$-BN samples \cite{CLAVIER2023108674, ZHEN2012261, Clavier2017}. These simulations were performed over 100 ps, with data collected every 25 fs (100 steps), to obtain the complete elastic constant matrix for each sample. Mechanical properties (Young's modulus, bulk modulus, and shear modulus) were then derived from the elastic constants obtained in the MD simulations \cite{Ray1984, Clavier2017}.
\\
5-nm thick Cu layers (z = 5 nm) were placed atop the $\rm \alpha$-BN films, as shown in Fig.\ref{fig:sample}-b. Heterostructures were equilibrated and relaxed for 20 ps for each structure. Later the structures were heated up by 50 K with a constant rate of 5 K/ps, and then equilibrated at the target temperature for 50 ps and the positions of atoms were printed at every 10 fs. This cycle was performed under NVT ensemble. At the end of the equilibration run, the heterostructures were heated again. This cycle was continued until the temperature reached 1000 K. We evaluated the barrier performance of the films using mean squared displacement (MSD). MSD measures how much an atom moves from its initial position: $\rm MSD(t) = \langle | r^{(i)}(t)-r^{(i)}(t=0) |^2 \rangle$. The slope of MSD is the diffusivity (D) of atoms, $\rm D(t) = \lim_{t\to\infty} MSD(t)/6t$. When the system is stable, diffusivity should ideally be zero (or close to zero due to the temperature) \cite{Zhou2016}. When we observe an abrupt change in the diffusivity of Cu atoms, this means $\rm \alpha$-BN cannot stop Cu atoms diffuse into the $\rm \alpha$-BN barrier and potentially to the substrate material. \\
\subsection{Tight Binding Model}
To estimate the electronic properties and dielectric constant of $\rm \alpha$-BN, we use a simple tight binding (TB) model based on a first nearest neighbors Slater-Koster approach \cite{Galvani2024}. The parameters of the model are fitted using data from the cubic, wurtzite, and single-layer hexagonal phases of BN at equilibrium and under 10\% isotropic dilation, with B–B and N–N hoppings approximated using B–N parameters. While this model captures the effects of local coordination and geometry via distance-dependent hoppings, it lacks additional energetic corrections which could be important here \cite{Galvani2024}.\\
We refer the reader to \cite{Galvani2024} and references therein for details and limitations, but for self-containedness, we outline the methodology here. Dielectric functions have been computed at $T=0 \text{ K}$, at the single particle, frozen atoms level, in the long wavelength limit ($\vb{q}\to 0$):\cite{Grosso2013solid}\footnote{In practice, we compute the complex dielectric function: $\epsilon(\vb{q}=\vb{0},\omega)=1+\frac{8\pi}{\Omega}\frac{\hbar^2}{{m_e}^2}\sum_{v,c}\frac{\abs{\mel{c}{\vec{e}\cdot\hat{\vb{p}}}{v}}^2}{\qty(E_c-E_v)^2} \qty[\frac{1}{E_c-E_v-\hbar\omega - i\eta}+\frac{1}{E_c-E_v+\hbar\omega + i\eta}]$ where $\eta\to0^+$ is a phenomenological broadening set here to $k_bT_{room}=26 \text{ meV}$. The expressions given in the maintext are the $\eta\to0^+$ limits.}
\begin{align}
\epsilon_1(\omega)=\text{Re}\qty[\epsilon(\omega)]&=1+\frac{16\pi}{\Omega}\frac{\hbar^2}{{m_e}^2}\sum_{v,c}\frac{\abs{\mel{c}{\vec{e}\cdot\hat{\vb{p}}}{v}}^2}{\qty(E_c-E_v)^2} \frac{E_c-E_v}{\qty(E_c-E_v)^2 - \qty(\hbar\omega)^2}  \label{eq:EpsilonReal}
\\
\epsilon_2(\omega)=\text{Im}\qty[\epsilon(\omega)]&=\frac{8\pi}{\Omega}\frac{\hbar^2}{{m_e}^2}\sum_{v,c}\frac{\abs{\mel{c}{\vec{e}\cdot\hat{\vb{p}}}{v}}^2}{\qty(E_c-E_v)^2} \delta\qty(E_c-E_v-\hbar\omega)     \label{eq:EpsilonImag}
\end{align}
in which $\Omega$ is the system's unit cell volume, $\hbar$ is the reduced Planck constant, $m_e$ is the electron mass, $\vec{e}$ is the polarisation of the electric field, $\omega$ its angular frequency, the $E_i$ and $\ket{i}$ are respectively the electronic eigenenergies and eigenstates of the TB Hamiltonian, with $v$ and $c$ running over the valence (occupied) and conduction (unoccupied) states, and $\vb{p}$ is the momentum operator. For clarity, we note that such a frozen atoms calculation excludes the vibrational contributions to the dielectric function.\\
Eigenstates and eigenenergies were obtained by exact diagonalization of the TB Hamiltonian at the $\vb{\Gamma}$ point. Following the MD, periodic boundary conditions were assumed in $(x,y)$ but not in $z$. Strictly, the unit cell height, and thus its volume, $\Omega$, is ill defined, as the films are two dimensional (and the unit cells thus contain arbitrary vacuum regions, as explained above) \cite{Cudazzo2011}. However, a natural choice of height is given here by the thickness of the films, through which a representative $\Omega$ is thus prescribed.\footnote{We use the difference $\Delta z$ in $z$ coordinate between the atom of highest and lowest $z$ for unit cell height, and the $(x,y)$ area $\mathcal{A}$ of the periodic cell for the effective volume: $\Omega \leftarrow \mathcal{A} \times \Delta z$.} Matrix elements of the momentum operator were obtained as in \cite{Galvani2024}, using a standard tight-binding approximation, which produces appropriate results for semiconductor systems \cite{Graf1995,Boykin2001electromagnetic,Boykin2001dielectric,Delerue2013nanostructures}.\\

It has been suggested \cite{Galvani2024} that $\rm \alpha$-BN can exhibit midgap states due to structural disorder and bond characteristics. These midgap states may increase the dielectric constant drastically since they allow low energy transitions. However, these states appear mostly localised in real space due the disordered nature of the samples in the samples. To visualise the formation of midgap states and their localisation effects, we calculated the electronic density of states (DOS)\footnote{The DOS are computed using the same eigenenergies as the ones used for the dielectric function calculation, and with the same lorentzian broadening of $26 \text{ meV}$.} and the inverse participation ratio (IPR) for all eigenstates of each film. The IPR provides a measurement of a state's localisation in real space. For an electronic eigenstate $\ket{\Psi}$, we adopt the definition
$
\text{IPR}(\ket{\Psi})=\sum_{\vb{n}}\qty[q_{\vb{n}}\qty(\ket{\Psi})]^2
$,
where the sum runs over atomic positions and $q_{\vb{n}}  \qty(\ket{\Psi})$ is the corresponding probability of presence of the electron at atomic site $\vb{n}$ (obtained by summing the modulus squared of the projection of $\ket{\Psi}$ on each of its orbitals). Heuristically, if $\ket{\Psi}$ describes a state localised on a single atomic site, say $\vb{m}$, $q_{\vb{n}} \approx \delta_{\vb{n},\vb{m}}$ and $\text{IPR}(\ket{\Psi}) \approx 1$. On the other hand, if $\ket{\Psi}$ describes a state delocalised over the whole system, the $q_{\vb{n}}$ will be of order $1/N_{\text{atoms}}$ and therefore $\text{IPR}(\ket{\Psi}) \to 0$.

\subsection{Experimental Studies}
\subsubsection{$\rm \alpha$-BN Film Growth}
To experimentally demonstrate the barrier properties of $\rm \alpha$-BN samples, we employed capacitively coupled plasma-chemical vapour deposition (CCP-CVD) to deposit $\rm \alpha$-BN films. The Si substrate was placed on a substrate holder in the deposition chamber. The substrate was heated at a rate of 10 ºC/min and reached growth temperatures between 350 and 400 ºC. Following a 20-minute annealing process, a borazine precursor ($\rm B_3H_6N_3$, Gelest) was introduced into the chamber at a rate of 0.05 sccm with a hydrogen flow of 20 sccm. During the growth process, the plasma operated at a power of 20 W. The growth time was adjusted between 20 and 80 minutes depending on the thickness of the $\rm \alpha$-BN film. After the growth was completed, the furnace was slowly cooled down to room temperature.

\subsubsection{Experimental Cu Diffusion Barrier Test}
 We assessed the Cu diffusion barrier properties of the grown $\rm \alpha$-BN films. For comparison with simulation results, 3-nm, 5-nm, and 7-nm thick $\rm \alpha$-BN films on Si substrate were prepared and a 50-nm thick Cu film was deposited on $\rm \alpha$-BN/Si by e-beam evapouration. Cu/$\rm \alpha$-BN/Si samples were annealed at the temperature range of 400 ºC to 600 ºC for 30 minutes using a vacuum furnace. Subsequently, these samples were immersed in a Cu etchant (49-1, Trensene) for 20 minutes to remove the Cu films, followed by a deionized water rinse.  to remove the excess etchant. Cu diffusion was evaluated through confirmation of the formation of copper silicide using X-ray diffraction (XRD) and scanning electron microscopy (SEM).
 
\section{Properties of $\rm \alpha$-BN Thin Films}
\subsection{Film Generation and Morphological Features} \label{ss:FilmsMorpho}
\begin{figure}[b!]
    \centering
    \includegraphics[width=0.6\columnwidth]{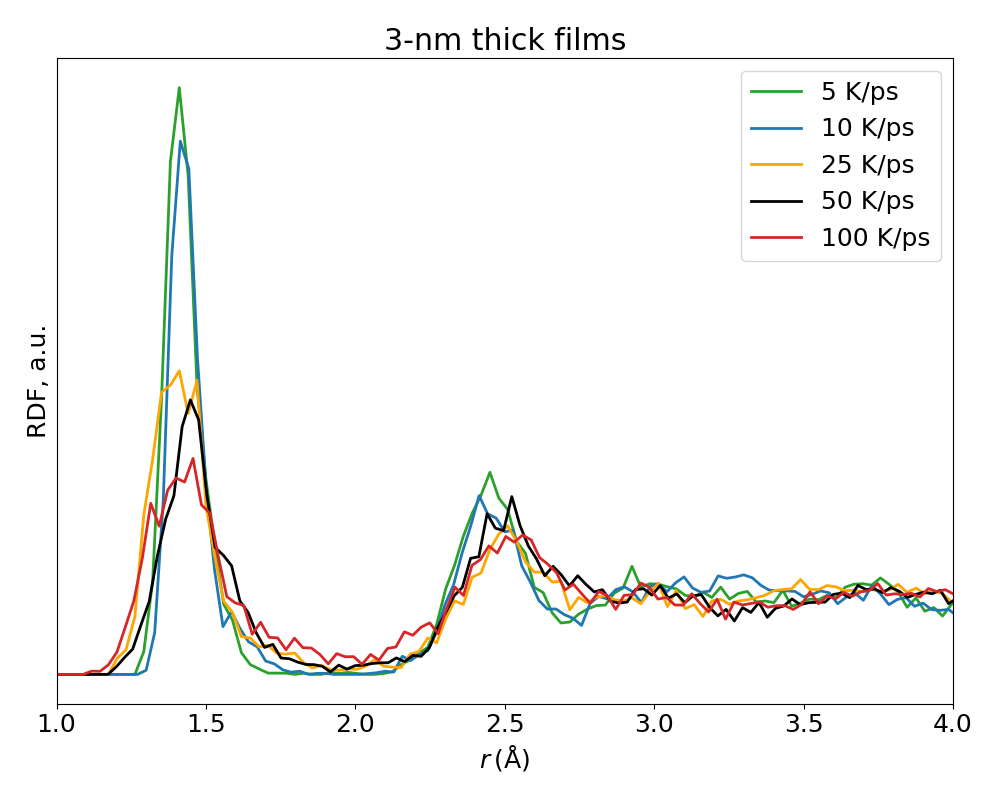}
    \caption{Radial distribution function of 3-nm thick $\rm \alpha$-BN samples with respect to cooling rate.}
    \label{fig:rdf}
\end{figure}
\begin{figure}[htb!]
    \centering
    \includegraphics[width=1.0\columnwidth]{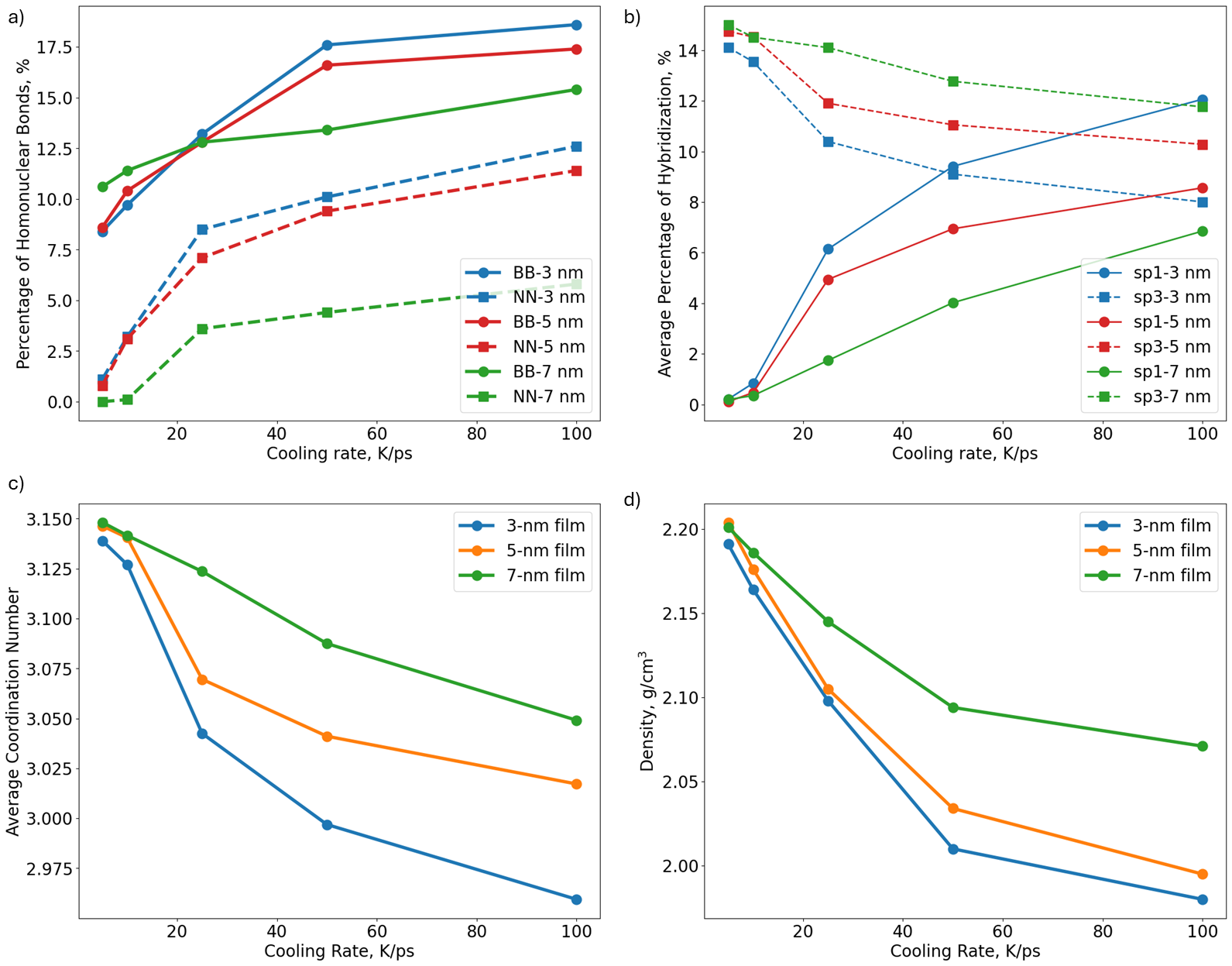}
    \caption{Variation of morphological features of $\rm \alpha$-BN thin films with respect to the applied cooling rate: (a) percentage of homonuclear bonds (B-B + N-N bonds), (b) percentage of $\rm sp^1$- and $\rm sp^3$-hybridised atoms, (c) average coordination number, and (d) density. Each data point represents an average over five samples.}
    \label{fig:morphology}
\end{figure}

After generating samples through the melt-quench protocol, we focused on key morphological features to identify the structural differences between fast- and slow-cooled structures. The morphology of $\rm \alpha$-BN, as with other disordered materials, critically influences material properties and device performance \cite{Glavin2016, Kaya2023,Lin2022, Galvani2024}. A thorough investigation of morphological features is therefore vital. To evaluate the  ``crystallinity vs. disorder'' of the samples, we first computed the radial distribution function (RDF), $\rho(r)$, of $\rm \alpha$-BN films as a function of cooling rate. Fig. \ref{fig:rdf} presents the average RDF of 3-nm thick films. As all thicknesses show the same trend, we only present the 3-nm thick films here, RDF of 5- and 7- nm thick films is presented in Fig. \ref{fig:rdf_films} in \ref{a:pe}. While all structures are characterised by a clear and recognisable first two peaks, no peak is identified for distances larger than 3 $\rm \AA$, indicating the absence of long-range order in the samples regardless of the cooling protocol employed and showing the amorphous character of the films is not affected by the cooling rate. The first peaks of the slow-cooled samples are close to 1.44-1.45 $\rm \AA$, which is the bond distance of $\rm sp^2$ B-N bonds, as can be seen in the close-up image of the first peak of the RDF presented in the inset of Fig. \ref{fig:rdf}. The broad peak in the RDF indicates a significant variation in bond lengths, reflecting a distribution between 1.35 - 1.55 $\rm \AA$, and is consistent with structural disorder.The peak broadens further and shifts away from 1.45 $\rm \AA$ with higher cooling rates, indicating more disordered structures.  Similar behaviour is observed in the 5-nm and 7-nm films as shown in Figs. \ref{fig:rdf_films} in \ref{a:pe}. With increasing film thickness, the first-neighbour peak in the RDF sharpens slightly, indicating improved short-range order. However, no additional peaks emerge beyond $\rm 3~\AA$, confirming the lack of long-range order even in the thickest films. This is consistent with the absence of medium-range features in the corresponding structure factors $S(q)$ (Fig. \ref{fig:sq_triangle} in \ref{a:pe}). $S(q)$ shows no features associated with medium-range order, and no first sharp diffraction peak is present \cite{Elliott1991}. While the intensity of the main peak in $S(q)$ increases modestly with thickness, no higher-order peaks appear, indicating that the structural network remains amorphous regardless of thickness. Fig. \ref{fig:morphology} presents how several morphological features change with the applied cooling rate. As shown in the figure, with higher cooling rates, the number of B-B bonds (which usually have a bond length of 1.6 - 1.8 $\rm \AA$) increased rapidly. In addition, we observed that $\rm sp^1$-hybridised atoms, since they are mostly located around voids and pinholes, have quite strongly varying bond distances, which also contributes to the broadening observed in the RDF. \\
Fig. \ref{fig:morphology} summarises trends between cooling rate and morphological features. Fig. \ref{fig:morphology}-a shows that higher cooling rates increase the proportion of both B-B and N-N bonds for all thicknesses. However, the increase in homonuclear bonds varies with film thickness. At any given cooling rate, thinner samples consistently contain more homonuclear bonds than thicker ones. For instance, the fraction of homonuclear bonds in 3-nm and 5-nm thick films generated at a cooling rate of 5 K/ps are approximately 10$\%$, rising to 30$\%$ at higher cooling rates, whereas for rapidly cooled (cooling rate $\textgreater$ 50 K/ps) the 7-nm thick sample, the proportion of homonuclear bonds remains below 20 $\%$. Fig. \ref{fig:morphology}-b shows how the coordination number of $\rm \alpha$-BN is influenced by applied cooling rate during film generation. As with the homonuclear fraction, the effect of cooling rate on local coordination is more evident in thinner films. While all structures cooled at 5 K/ps have no $\rm sp^1$-hybridised atoms (coordination number of 2), their proportion rapidly increases to 10$\%$ in 3-nm thick films at higher cooling rates. Similarly, for 5- and 7-nm thick films, the fractions of $\rm sp^1$-hybridised atoms become $\rm \sim 8\%$ and $\rm \sim 6\%$, respectively. Moreover, the number of $\rm sp^3$-hybridised atoms (coordination number of 4) drops rapidly at all thicknesses as cooling rate increases. These changes in hybridisation are reflected in the average coordination number (Fig. \ref{fig:morphology}-c), larger cooling rates lead to a rapid decrease in average coordination at all thicknesses, most notably in the 3-nm films. While $\rm sp^1$- and $\rm sp^3$-hybridised fractions show opposing trends with the cooling rate, the average coordination number reflects the overall change in the local atomic environment. With larger cooling rates, the average coordination number of samples at all thicknesses drops quickly, with the largest change observed for 3 nm thick films. A parallel trend is also observed in mass density of the samples. As for the dielectric function, we estimated the film volume as $\mathcal{A}\times\Delta z$ using the difference in $z$-coordinate $\Delta z$ between the atoms of highest and lowest $z$ and the $(x,y)$ area $\mathcal{A}$ of the periodic cell, as depicted in Fig. \ref{fig:morphology}-d. as depicted in Fig. \ref{fig:morphology}-d. The decrease in coordination number, as cooling rates increase, indicates increased interatomic separation, which lowers the density. While the density of films is $\rm \sim 2.21$ $\rm g/cm^3$ for samples cooled at 5 K/ps, the density drops below 2 $\rm g/cm^3$ for 3-nm and 5-nm thick films at higher cooling rate. Meanwhile, even the 7-nm thick samples generated with the largest cooling rate still have an average density of $\rm \sim 2.07$ $\rm g/cm^3$. The significant reduction in density, together with the increase in $\rm sp^1$-hybridised atoms suggests that fast-cooled films have more porous structures.Fig. \ref{fig:pores}, presented in \ref{a:pe}, shows voids in fast-cooled (50 K/ps and 100 K/ps) 3-nm $\rm \alpha$-BN films. Around these defects we also observe $\rm sp^1$-hybridised atoms, emphasise the relationship between voids and $\rm sp^1$ environments. \\
 With increasing thickness, the reduction in homonuclear bonds and $\rm sp^1$-hybridised atoms shown in Fig. \ref{fig:morphology} indicates further structural relaxation in thicker films. This improvement in structural order is most pronounced in the 7-nm films, whereas the 3-nm and 5-nm films show smaller changes. The behaviour in the 7-nm films suggests a greater capacity to accommodate strain and to attain a more ordered atomic arrangement as thickness increases. Similar behaviour has been reported for amorphous systems \cite{Agirseven2022, Moghadam2016, Trofimov2006}, where larger volumes facilitate relaxation processes and reduce disorder. Although thickness favours relaxation, larger thicknesses during growth can facilitate ring alignment and the emergence of turbostratic or nanocrystalline h-BN \cite{Kim2023}. This competing crystallisation pathway can modify dielectric response and diffusion-barrier performance.\\
\subsection{Mechanical Properties}
Strong mechanical performance is critical for diffusion barriers because low strength and a low Young’s modulus complicate integration. A robust mechanical response indicates that the film resists deformation under stress, which is crucial for structural integrity during processing and operation \cite{Hong2020, framil2019influence, wang2004mechanical}. As with other properties, the mechanical behaviour of $\rm \alpha$-BN films is strongly influenced by morphology and fabrication conditions. Here, we investigated the Young’s, bulk, and shear moduli of $\rm \alpha$-BN films by computing the elastic constants as a function of film thickness and cooling rate.
\begin{figure}[ht!]
    \centering
    \includegraphics[width=0.75\columnwidth]{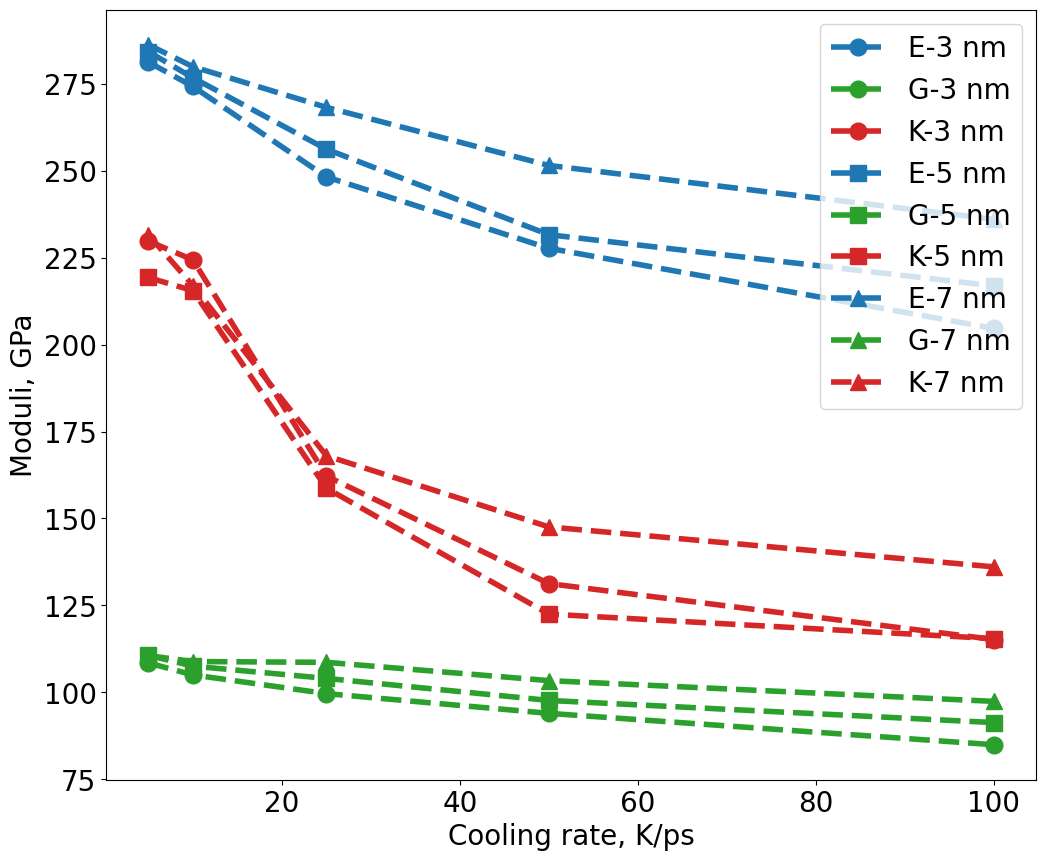}
    \caption{Young's modulus (E), shear modulus (G) and bulk modulus (K)) of $\rm \alpha$-BN films at different thicknesses and generated with different cooling rates.}
    \label{fig:mechanical}
\end{figure}
The elastic constants for each film were calculated as described in the Methods section. Young’s modulus and the other mechanical moduli with respect to thickness and applied cooling rate are presented in Fig. \ref{fig:mechanical}. The resulting mechanical properties exhibit trends that parallel the morphological features. Films generated at higher cooling rates, as presented in Fig. \ref{fig:morphology}, contain reduced fractions of $\rm sp^3$- and $\rm sp^2$-hybridised atoms, a higher proportion of homonuclear bonds, lower density, and increased porosity (Fig. \ref{fig:pores} in \ref{a:pe}). These films display a $\rm \sim 20\%$ lower Young’s modulus than more slowly cooled samples. The other moduli also decrease significantly with increasing cooling rate. The high Young’s modulus of the slow-cooled samples can be attributed to the higher fractions of $\rm sp^2$- and $\rm sp^3$-hybridised atoms and to the absence of the pores and voids observed in the fast-cooled samples (Fig. \ref{fig:pores}). The positive impact of $\rm sp^3$-hybridised environments on stiffness is well established \cite{Kaya2023,kaya2023impact,Jana2019}. Young’s modulus is particularly critical for diffusion barriers because a higher value indicates greater stiffness and resistance to deformation. This is essential to ensure that the barrier maintains structural integrity during the high-stress conditions typical of semiconductor processing and operation. A higher Young’s modulus in the slow-cooled films suggests that these barriers would better resist compressive and tensile stresses, minimising risks such as cracking or delamination that could compromise diffusion-barrier performance. In contrast, the lower Young’s modulus observed in the fast-cooled films may render them more susceptible to deformation under stress, which could lead to voids and defects during device fabrication and operation. However, a very high Young’s modulus in the barrier layer can be problematic if there is a mismatch between the thermal expansion of the barrier and adjacent layers, since a stiffer structure can concentrate stress and promote deformation.\\
Although not shown in the figure, Poisson's ratio exhibits a weak dependence on the applied cooling rate. For films cooled at 5 K/ps and 10 K/ps, Poisson's ratio remains approximately constant at 0.28–0.30 across thicknesses, indicating a consistent balance between lateral expansion and contraction under stress. In contrast, faster-cooled samples exhibit a lower Poisson's ratio (0.19–0.22), indicating reduced lateral compliance. This subtle change may affect the barrier's ability to accommodate mechanical strains and its performance during thermal cycling and integration processes.
Additionally, the mechanical moduli exhibit a clear dependence on film thickness, thinner films consistently exhibit lower values across all cooling rates. This behaviour is consistent with the morphological differences discussed in \ref{ss:FilmsMorpho}. Specifically, thinner films contain a higher fraction of $\rm sp^1$-hybridised atoms, a larger proportion of homonuclear bonding, lower mass density, and increased porosity (see Fig. \ref{fig:morphology} and \ref{a:pe}). These structural features reflect a higher degree of disorder and reduced network connectivity, resulting in a softer and mechanically less robust amorphous matrix. In contrast, enhanced structural relaxation in thicker films allows them to accommodate local strain and form a more rigid network, which increases stiffness.
\subsection{Dielectric Properties of $\rm \alpha$-BN Samples}
Replacing the current metal barriers with low dielectric constant material is desirable because it increases the the cross-sectional area available for low-resistivity metals (such as Cu) and reduces the overall line resistivity. $\rm \alpha$-BN has already been studied for this use \cite{Galvani2024,Lin2022,Glavin2016}, however both experimental and theoretical results show a wide range dielectric constant values, strongly suggesting that this property, as well as mechanical and structural properties, depends on the morphology \cite{SattariEsfahlan2025}. \\
In this section, we present the real and imaginary parts of the dielectric functions of the films along the z-direction, which corresponds to the perpendicular to the plane of the thin film. The figure of merit we focus on in this section is the static dielectric constant $\rm \epsilon_1(\omega=0)$ of the films. \\ 
\begin{figure}[htb!]
    \centering
    \includegraphics[width=\columnwidth]{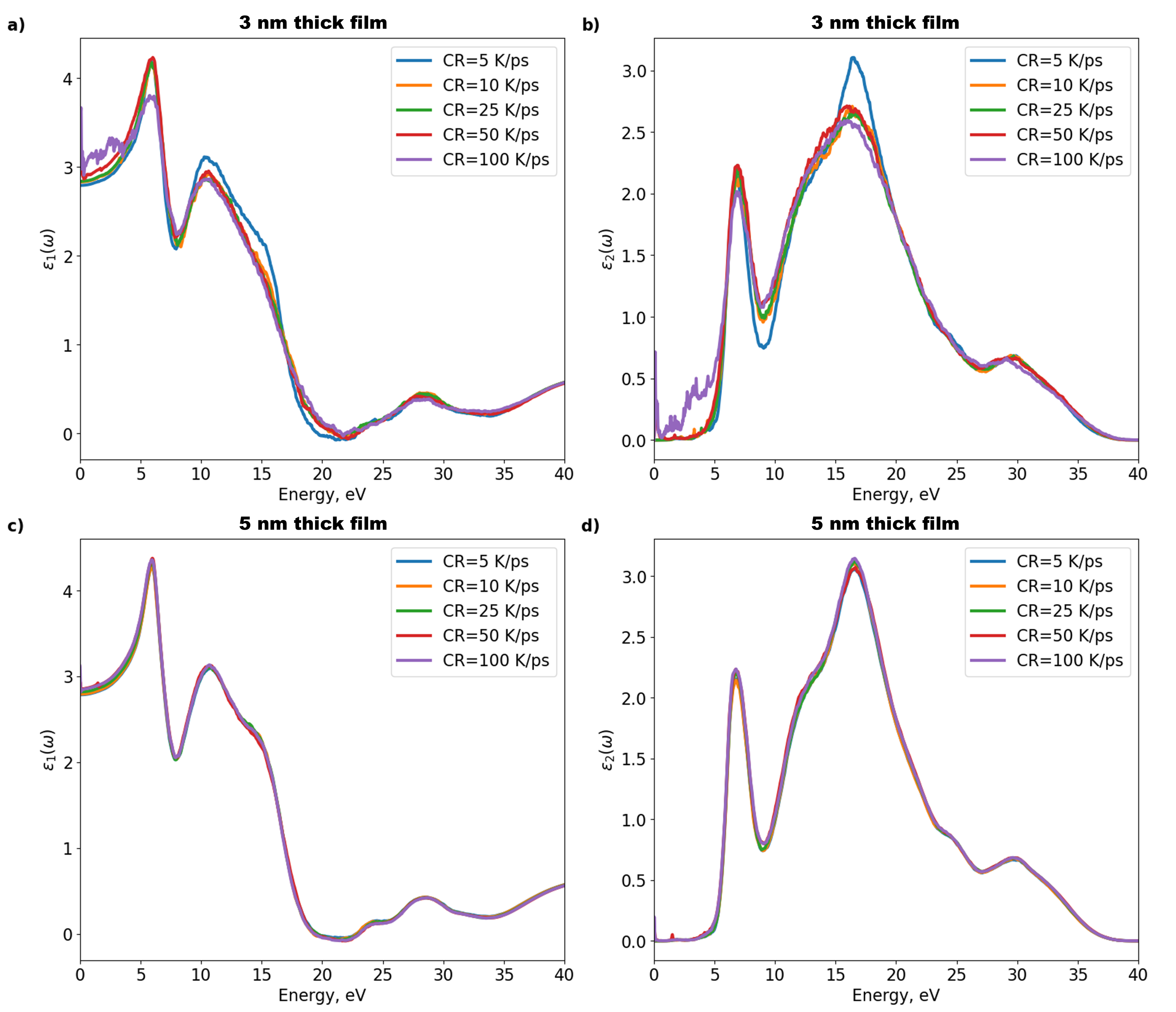}
    \caption{Average real (a,c) and imaginary (b,d) parts of the dielectric function of 3-nm and 5-nm thick $\rm \alpha$-BN films along the z direction as a function of cooling rate.}
    \label{fig:dielectric}
\end{figure}
Fig. \ref{fig:dielectric} shows the z-component of the real and imaginary parts of the dielectric function for 3-nm (a, c) and 5-nm thick (b, d) films.\footnote{Electronic properties for 7-nm films were not evaluated due to computational limitations.} The results are quite similar at both thicknesses. While slow-cooled samples (CR$\rm <$25 K/ps) at both thicknesses exhibited relatively low dielectric constants, around 2.8 (Fig. \ref{fig:dielectric}-a), the dielectric functions of the two fastest-cooled samples displayed a tendency to diverge at low frequencies due to the absence of an electronic gap. Although there seems to be a relationship between cooling rates and dielectric constants, the values are too close to establish a definitive trend of the variability of the dielectric properties upon increasing the film thickness.\\
To further analyse the obtained results, it is of interest to consider the energy and ``strengths" $D_{cv}^{\vec{e}} = \frac{\abs{\mel{c}{\vec{e}\cdot\hat{\vb{p}}}{v}}^2}{\qty(E_c-E_v)^2}$ of electronic transitions in the sample.\footnote{Up to constant prefactors, the $\frac{\abs{\mel{c}{\vec{e}\cdot\hat{\vb{p}}}{v}}^2}{\qty(E_c-E_v)^2}$ are essentially oscillator strengths along $\vec{e}$ divided by transition energies.} Indeed, $\rm \epsilon_1(\omega=0)$, up to constants, can be seen as the sum of these strengths over all transitions the system, inversely weighted by the transition energies (see equation \ref{eq:EpsilonReal} and, e.g. \cite{Galvani2024}). Conveniently, the imaginary part of the dielectric function resolves in energy the density and strengths $D_{cv}^{\vec{e}}$ of electronic transitions (per equation \ref{eq:EpsilonImag}, it can be interpreted as a density of states for electronic transitions, with each transition weighted by the corresponding strength $D_{cv}^{\vec{e}}$ \cite{Grosso2013solid}). For fast-cooled samples, the imaginary part of the dielectric function displays peaks at low energies, signaling the existence of low-energy transitions (of non-vanishing strengths), which are absent in the slowly-cooled samples.
\begin{figure}[htb!]
    \centering
    \includegraphics[width=\columnwidth]{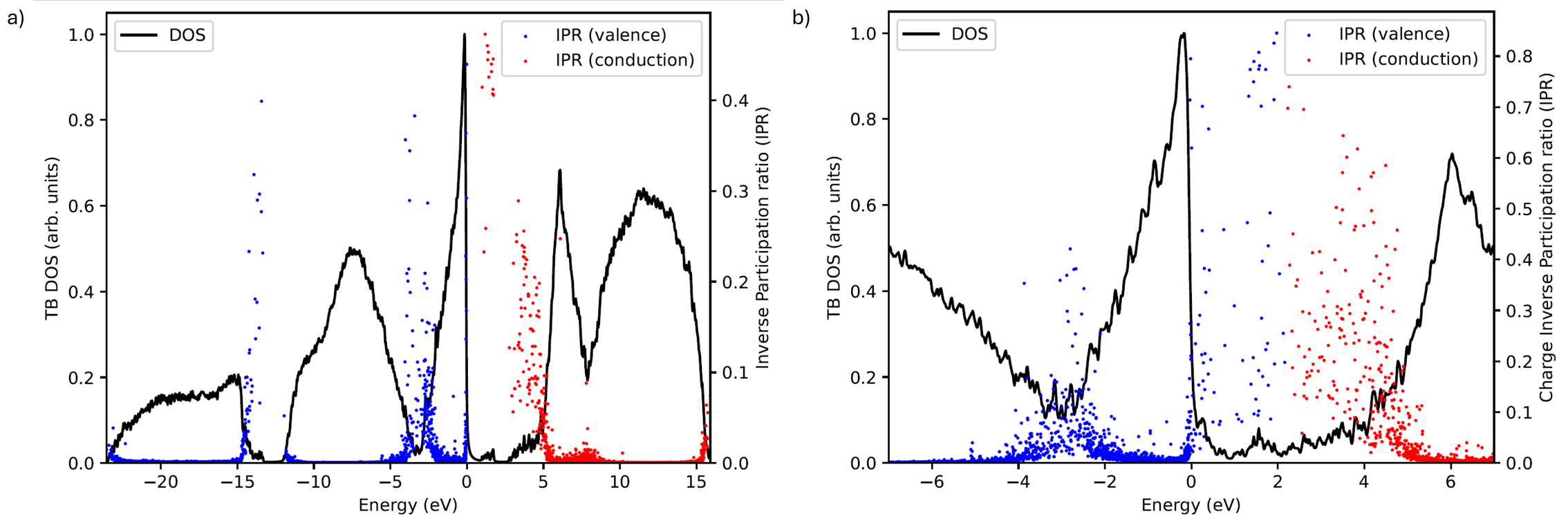}
    \caption{Electronic DOS and inverse participation ratio of 3-nm thick $\rm \alpha$-BN samples generated with cooling rates of 5 K/ps (a) and 100 K/ps (b).}
    \label{fig:DOS}
\end{figure}
To get further insight into these transitions and the dielectric constant behaviour, we examine the DOS and IPR of the films. In Fig. \ref{fig:DOS}, we present, for simplicity, the 3-nm thick films generated with cooling rates of 5 K/ps and 100 K/ps. Both films can be seen to exhibit mid-gap states. However, while the slow-cooled sample (\ref{fig:DOS}-a) shows a nominal (including mig-gap states) electronic gap of $\rm \sim 1.2$ eV, the fast-cooled sample (\ref{fig:DOS}-b) displays an effectively closed gap, allowing for the aforementioned low-energy transitions. In the slow-cooled sample, it appears that most midgap states are unoccupied (red dots in Fig. (\ref{fig:DOS}-a)), while the fast-cooled sample exhibits occupied mid-gap states. In both cases, mid-gap states show IPRs$>$0, indicating real-space localisation, which typically suppresses momentum matrix elements between them (numerators in equations \ref{eq:EpsilonReal} and \ref{eq:EpsilonImag}) and DC conductivity, lowering (but here not eliminating) their impact on the dielectric constant.
At this point, we should provide qualitative remarks regarding the expected impacts of structure on the dielectric constant of aBN systems. DFT calculations in \cite{Galvani2024} suggest that the dielectric constant of bulk $\rm \alpha$-BN is increased by the presence of $\rm sp^1$- and $\rm sp^3$- coordinated atoms (in $\rm sp^2$-dominated samples) and local structural disorder, which can be related to the presence of homonuclear (B-B and N-N) bonds. Lin et al. \cite{Lin2022} also point towards the detrimental effects of Boron clusters.\footnote{Although the 3 nm thick sample cooled at 25 K/ps had a significant number of N-N bonds, we did not observe any large N or B clusters. However, we observed some B and N clusters (B-B or N-N bonds with a size of 4-12 atoms).} Similarly, Glavin et al. \cite{Glavin2016} indicated that B-B bonds may promote the formation of mid-gap states or conductive pathways in thin films. These features, which we expect in fast-cooled films, as shown in Fig. \ref{fig:morphology}, may thus contribute to the formation of midgap states and increase the dielectric constant in these films.\\
As can be inferred from the presence of the unit cell volume in the denominator of equation \ref{eq:EpsilonReal}, with all other parameters unchanged, the dielectric constant is also expected to increase with density. Our results (Fig. \ref{fig:morphology} d.) show that density decreases in fast-cooled films compared to slower cooling rates. This suggests an opposite trend to the one discussed above: from this effect \emph{alone}, we would expect a decrease of $\epsilon_1(\omega=0)$ on the order of $\sim 10\%$ from the 3-nm and 5-nm films cooled at $100 \text{ K/ps}$ compared to those cooled at $5 \text{ K/ps}$ (as per the changes in density reported in Fig. \ref{fig:morphology} d.). As can be seen in Fig. \ref{fig:dielectric}, we do not observe such a reduction, which hints at the importance of the aforementioned structure-induced changes to the electronic structure.\\
Thus, while fast cooling rates favor a lower density (and thus volume), which, if unit cell polarizability was unchanged, would lower the dielectric constant, it also favors morphological features (homonuclear bonds, $\rm sp^1$- and $\rm sp^3$- coordinated atoms, etc.) which can increase the polarizability per unit cell. \\
It is possible that, within our simple tight-binding model, these effects largely cancel each other, leading only to a very weak dependence of dielectric constant on cooling rates. As stated above, we still observe a very weak effect for both thicknesses suggesting that $\epsilon_1(\omega=0)$ could increase with faster cooling rates. Considering the simplicity of the tight-binding model, we cannot however confidently establish a definite trend. Within this caveat, we nevertheless note that thinner (3 nm) films appear to display more significant variations of their dielectric functions (and constants) with cooling rate. This appears consistent with the findings of section \ref{ss:FilmsMorpho}: the thinner the film, the stronger the effects of cooling rate variations in inducing homonuclear bonds and $\rm sp^1$- and $\rm sp^3$- coordinated atoms. \\
From Fig. \ref{fig:DOS} we observe that there appears to be a gap of $\sim 5 \text{ eV}$ between unoccupied and occupied states with low IPR (thus expected to be more delocalised with respect to the midgap states). Observation of $\epsilon_2(\omega)$, in Fig. \ref{fig:dielectric}, suggests that this is the case in most samples, with these states giving rise to the first main peak (at the single-particle level), albeit lower energy contributions from the (significantly less numerous) mid-gap states can also be observed. Here again, the strongest change with cooling rate is observed for the 3-nm films. For this thickness, in the case of the $100 \text{ K/ps}$ cooling rate, we observe a significantly higher response at low ($< 5 \text{ eV}$) frequencies (Fig \ref{fig:dielectric} b.), which correspondingly increases the dielectric constant (Fig \ref{fig:dielectric} a.). This effect is significantly less pronounced in the case of 5-nm films, further hinting at the fact that properties of thicker films may be less dependent on cooling rate.
\section{Theoretical Investigation of $\rm \alpha$-BN Diffusion Barriers}
\begin{figure}[htb!]
    \centering
    \includegraphics[width=0.7\columnwidth]{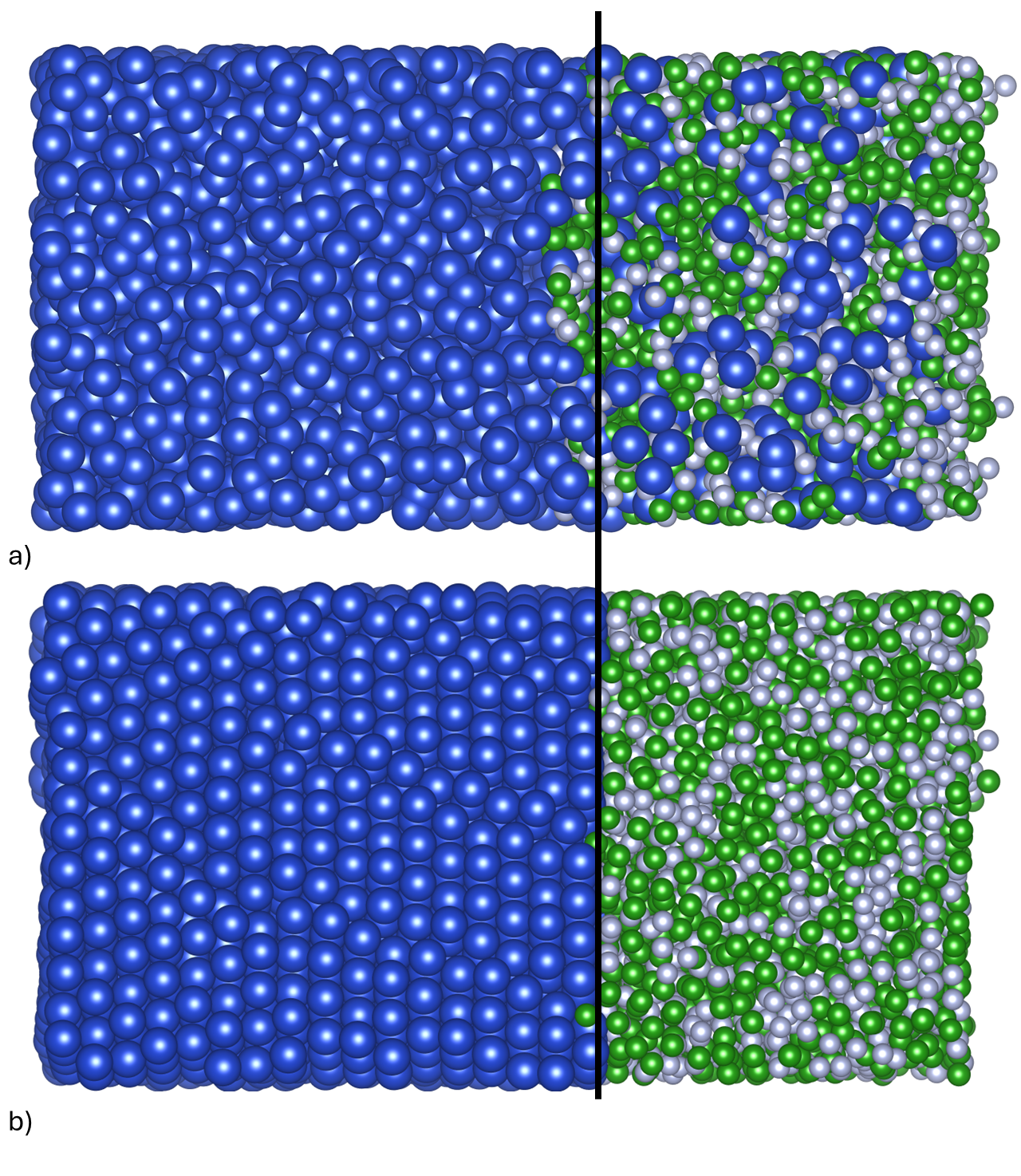}
    \caption{Examples of a failed and successful 3-nm thick $\rm \alpha$-BN diffusion barriers against Cu diffusion after the annealing process until 1000 K. While the failed sample was generated using a cooling rate of 100 K/ps (a), a rate of 5 K/ps (b) was applied for the successful one. Black solid line shows the approximate position of the interface before the annealing process.}
    \label{fig:void}
\end{figure}

 The effectiveness of $\rm \alpha$-BN films as copper diffusion barriers hinges on their ability to resist Cu migration into the barrier structure. This section investigates how the morphology of $\rm \alpha$-BN, controlled by cooling protocols and film thickness, influences its diffusion resistance. By studying the interaction of Cu atoms with $\rm \alpha$-BN at the atomic scale, we aim to elucidate the structural features critical for preventing copper penetration.\\
Molecular dynamics (MD) simulations, as described in Section 2.1, were used to model $\rm \alpha$-BN/Cu heterostructures. $\rm \alpha$-BN films were generated under varying cooling rates to capture a range of disordered morphologies. An annealing protocol, also detailed in Section 2.1, incrementally raised the temperature from 300 K to 1000 K to mimic the experiments presented in Section 5. During this process, the diffusivity of Cu atoms within the $\rm \alpha$-BN layer was monitored using MSD calculations, and atomic trajectories were analysed to identify mechanisms and pathways enabling copper diffusion. These simulations provide insights into the interplay between $\rm \alpha$-BN morphology and its effectiveness as a diffusion barrier. \\
\begin{figure}[htb!]
    \centering
    \includegraphics[width=\columnwidth]{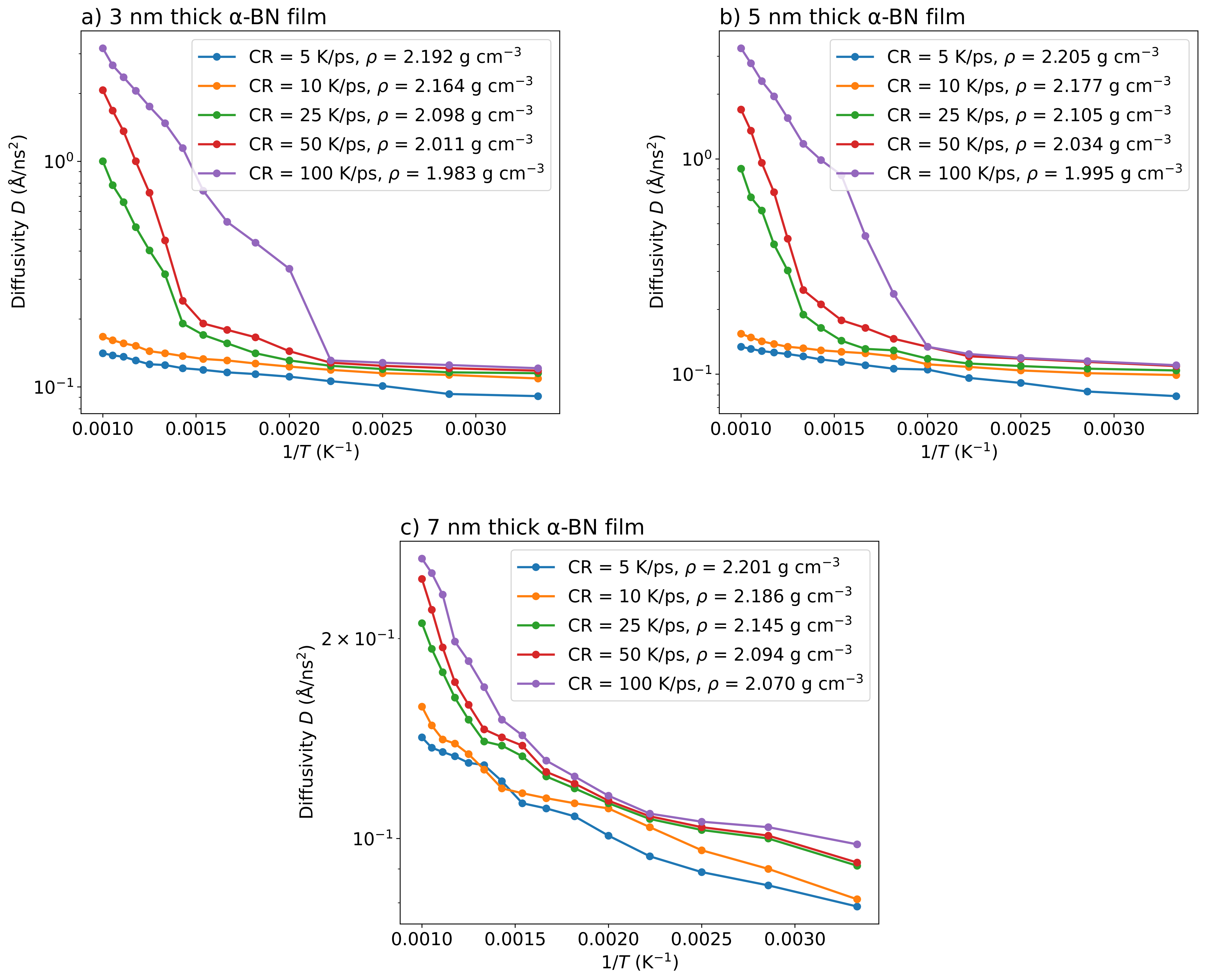}
    \caption{The performance of 3-nm (a), 5-nm (b), and 7-nm (c) $\rm \alpha$-BN films as diffusion barriers by evaluating the diffusivity, log $D$, of Cu atoms. Systematic deviations from the linear Arrhenius behaviour in log $D$ is considered as Cu diffusing into the $\rm \alpha$-BN film. Densities are for the $\rm \alpha$-BN layer at 300 K.}
    \label{fig:msd-cu}
\end{figure}
Fig.~\ref{fig:msd-cu} presents the Cu diffusivity, $D$, as a function of inverse temperature for 3~nm (a), 5~nm (b), and 7~nm (c) films to identify when the films lose their stability and protective abilities. $D$ values are non-zero since the atoms move slightly due to temperature. With higher temperature, we observe a linear increase in log $D$ of Cu as expected. While behaviour up to 450 K is linear-like at all thickness and cooling rates, we start to observe some deviations from Arrhenius behaviour with increasing temperature. Up to this temperature, we assume that the $\rm \alpha$-BN barriers are structurally stable and do not allow Cu atoms diffuse into the structure. However, with higher temperatures, amorphous films with higher bond and coordination defect concentration (B-B/N-N bonds, $\rm sp^1$-hybridised atoms) and lower density (which indicates larger free volume and nanovoids) may suffer from structural changes \cite{RosaJunior2019} with higher temperatures and diffusion of Cu atoms. The deviation from linear Arrhenius behaviour of Cu diffusion indicates a change in the barrier integrity and film morphology. Fig. \ref{fig:msd-cu} and \ref{fig:Dvsrho} in \ref{a:Dvsdensity} show that the barrier performance is strictly about the morphology of films. Beyond the temperature-dependent trends in Fig. \ref{fig:msd-cu}, a clear structure–property link is evident between the film mass density at 300 K and copper mobility. Within each thickness subset, log $D$ decreases monotonically as density increases, consistent with reduced free volume and fewer nanovoids limiting diffusion paths as shown in Fig. \ref{fig:Dvsrho} in \ref{a:Dvsdensity}. The effect is most pronounced in thinner films, and weakens as thickness increases, which aligns with the coordination and density trends in Fig. \ref{fig:morphology}-d.  \\
Slow-cooled (5 and 10 K/ps) 3 nm thick films have lower fractions of homonuclear bonds and $\rm sp^1$-hybridised atoms and have higher mass density, as discussed in Section \ref{ss:FilmsMorpho} and Fig. \ref{fig:morphology}. In addition, Fig.~\ref{fig:pe} in \ref{a:pe} shows that these films are more relaxed and stable than those generated at higher cooling rates, consistent with their higher mass density and the correspondingly lower free volume and void population compared to fast-cooled samples (CR $>$ 25 K/ps). As shown in Fig.~\ref{fig:msd-cu}, Cu atoms adjacent to these films exhibit significantly low diffusivity and follow an Arrhenius-like trend up to high temperatures ($>$800 K), with only minor deviations beyond this point. This suggests that the structure remains largely stable throughout the annealing process and continues to prevent or strongly limit Cu diffusion. However, fast-cooled samples ($>$25 K/ps) show a different behaviour. While films at 100 K/ps only manage to stop Cu atoms until 450 K, the ones at 50 K/ps and 25 K/ps manage to show linear-like behaviour until 600 to 700 K as shown in Fig. \ref{fig:msd-cu}. As discussed in Section~\ref{ss:FilmsMorpho}, these films contain higher fraction of bonding and coordination defects—namely homonuclear bonds and $\rm sp^1$-hybridised atoms—which lower their mass density and increase free volume. With rising temperature and Cu diffusion, existing nanovoids grow and these defect-rich regions lose stability, opening pathways for Cu transport. As Cu penetrates the $\rm \alpha$-BN layer near the interface, B and N atoms also migrate into the Cu layer, further accelerating void growth and ultimately leading to complete loss of barrier integrity. The difference in structural response and barrier performance can be directly observed in Fig.~\ref{fig:void}. The film generated at 100 K/ps (Fig.~\ref{fig:void}a) fails to block Cu, allowing full penetration into the $\rm \alpha$-BN layer, whereas the film at 5~K/ps (Fig.~\ref{fig:void}b) successfully prevents Cu diffusion.  \\
Similar morphology-barrier performance relationship is observed for 5 nm thick films with slightly more resilience. Films generated at 5 and 10 K/ps contain low fraction of $\rm sp^1$ and homonuclear bonds, have higher density, and less free volumes. These morphological features lead to more stable barrier performance that manage to stop Cu diffusion until high temperatures and Cu diffisuvity only exhibit a minimal deviation from linear Arrhenius behaviour as shown in Fig. \ref{fig:msd-cu} and \ref{fig:Dvsrho} in \ref{a:Dvsdensity}. However fast cooled samples generated at CR above 25 K/ps contained less stable regions with $\rm sp^1$-hybridised atoms and homonuclear bonds. This result in a loss of barrier integrity and quick increase in $D$ of Cu atoms at 500 K for films generated at 100 K/ps and 700 - 800 K for films generated at 25 and 50 K/ps. Similar to 3 nm films, as temperature increases further, defect-rich regions soften and voids grow, allowing gradual Cu penetration and some B and N migration into the Cu layer. Thus, while the same mechanism seen in the 3 nm case operates here, the increased thickness raises the temperature window over which the barrier can maintain its protective function.\\
At 7 nm, $\rm \alpha$-BN barriers exhibit a different response from the thinner films. As shown in Fig. \ref{fig:msd-cu}-c, Cu atoms have remarkably lower diffusivity values in 7 nm systems across full temperature range. While Cu atoms show linear Arrhenius-like behaviour until 600 K, films generated at higher CR started to slightly deviate from the linear behaviour, indicating a small Cu diffusion into $\rm \alpha$-BN films and small possible change in the morphology of the films. This different behaviour can be explained the morphology of the films presented in Fig. \ref{fig:morphology}. Figures~\ref{fig:morphology} and \ref{fig:pe} show that 7 nm films consistently contain fewer homonuclear bonds and a lower fraction of $\rm sp^1$-hybridised atoms than the thinner samples, together with higher coordination numbers, higher mass density, and correspondingly reduced free volume and voids. Even the rapidly quenched films start from a more relaxed energy state than 3 or 5 nm counterparts, which prevents early softening of weak regions during heating. These combined features demonstrate that the thicker films dilute the influence of interfacial defects and stabilise the amorphous network. 
These results show that the diffusion-barrier response of $\rm \alpha$-BN is governed by morphology rather than thickness alone. Slow-cooled 3–5 nm films (5–10 K/ps) maintain low Cu mobility up to 1000 K, whereas rapidly quenched networks ($>$50 K/ps) with higher coordination defects and voids lose integrity by 500–600 K. Within each thickness, higher density together with reduced homonuclear bonding and a lower fraction of sp$^{1}$-hybridised atoms lead to lower Cu diffusion. Increasing thickness mitigates, but does not remove, the sensitivity to the morphology. Practically, ultrathin $\rm \alpha$-BN can match the performance of thicker layers when the growth route yields a dense, well-connected network with minimal defects and free volume.

\section{Experimental Investigation of $\rm \alpha$-BN Diffusion Barriers}
\begin{figure}[htb!]
    \centering
    \includegraphics[width=1.0\columnwidth]{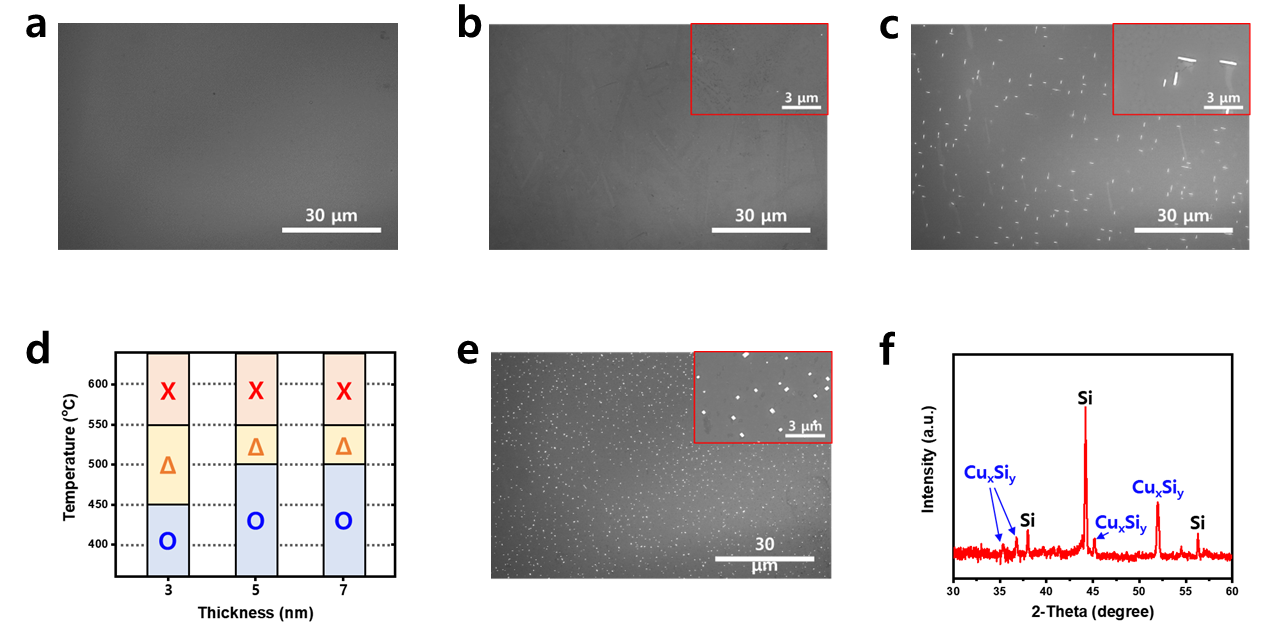}
    \caption{Experimental result of Cu diffusion barrier test. (a-c) SEM images of 3-nm thick $\rm \alpha$-BN/Si surface after 400 ºC (a), 500 ºC (b), and 600 ºC (c) annealing and removing residual Cu films with Cu etchants. (d) Evaluation of Cu diffusion barrier properties of 3-nm, 5-nm, and 7-nm thick $\rm \alpha$-BN/Si after 400 - 600 ºC annealing process. Blue O means $\rm \alpha$-BN film successfully blocked Cu diffusion. Yellow $\rm Delta$ means there is small copper silicide nucleation through the defect of $\rm \alpha$-BN films. Red X means $\rm \alpha$-BN film failed to block Cu diffusion and copper silicide crystals are formed. (e) SEM image of Si surface without $\rm \alpha$-BN barrier films after 400 ºC annealing and removing residual Cu films with Cu etchants. (f) XRD spectrum of Si surface with copper silicide crystals without $\rm \alpha$-BN barrier films that shows the formation of copper silicide crystals.}
    \label{fig:experiments}
\end{figure}

After we showed the potential of $\rm \alpha$-BN films using theoretical models, we conducted an experimental investigation on the diffusion barrier performance of $\rm \alpha$-BN films with the same thicknesses we performed our simulations. We fabricated Cu/$\rm \alpha$-BN/Si structures and annealed them at 400 - 600 ℃ for 30 minutes. After annealing at 400 ℃, a Si substrate coated with a 3-nm thick $\rm \alpha$-BN film did not form any copper silicide (Fig. \ref{fig:experiments}-a), while a Si substrate not covered with $\rm \alpha$-BN formed a substantial amount of copper silicide (Fig. \ref{fig:experiments}-e). After annealing 3-nm thick $\rm \alpha$-BN/Si at 500 ℃, very small copper silicide nucleations were observed on the Si surface (Fig. \ref{fig:experiments}-b), and the formation of copper silicide crystals was confirmed after annealing at 600°C (Fig. \ref{fig:experiments}-d). Fig. \ref{fig:experiments}-c provides a brief summary of Cu diffusion barrier properties of 3-nm, 5-nm, and 7-nm thick $\rm \alpha$-BN films after the 400 - 600 ℃ annealing process. The observation of crystallisation by SEM is fine but not sensitive enough to determine Cu diffusion in the sub-silicide formation regime - that is - you could have Cu in the Si but SEM resolution is insufficient to detect or within the solubility limit of Cu in Si (Fig. \ref{fig:sem} in \ref{a:sem}). The characteristics such as density or existence of defects of experimentally synthesized $\rm \alpha$-BN films differ from the simulated $\rm \alpha$-BN films and do not exhibit the same barrier performance as the simulations. However, there is a tendency for the barrier property to improve as the thickness of the $\rm \alpha$-BN film increases. 
\section{Conclusions}
We have investigated the diffusion barrier properties of ultrathin $\rm \alpha$-BN films against copper penetration by molecular dynamics and complemented the simulations with experimental data. Simulations show that Cu diffusivity is strongly suppressed in dense, slowly quenched films with low concentrations of B–B and N–N bonds, reduced $ \rm sp^1 $ content, and high atomic packing. These structures remain effective up to high temperatures, while rapidly quenched, defect-rich networks permit Cu diffusion at lower temperatures due to increased homonuclear bonding and reduced structural cohesion. Increasing the film thickness enhances performance by enabling structural relaxation and lowering defect concentrations. These trends are consistent with experimental measurements on PECVD-deposited 3-, 5-, and 7-nm films, which show barrier failure temperatures that increase with thickness. A central experimental challenge, however, is to preserve a controlled amorphous structure in BN films as their thickness increases, since B-N rings intrinsically tend to align and nucleate turbostratic or nanocrystalline h-BN domains \cite{Kim2023}. At small thicknesses, limited adatom mobility and kinetic constraints favour the stabilisation of a disordered network. However, as growth proceeds, enhanced surface and bulk diffusion, stress relaxation, cumulative plasma energy input, and local deviations from the ideal B:N stoichiometry progressively promote the emergence of ordered regions. This structural evolution not only undermines the targeted amorphousness but also modifies critical functional properties – including dielectric constant, density, and diffusion-barrier performance – thus making crystallisation control a key bottleneck for scaling $\rm \alpha$-BN films in interconnect and dielectric applications. To advance in this direction, it would be valuable to complement experiments with computational insights, for instance on the dependence of the dielectric constant on growth rate \cite{Lin2022}, or on the roles of growth temperature and oxygen incorporation \cite{chen2023tailoring}. Further computational studies of $\rm \alpha$-BN with larger thickness are also needed to explore (and find ways to mitigate) the effect of crystallisation observed in experiments, as it could jeopardise reaching the upper performance and integrability of this innovative advanced material into mainstream interconnect technologies. Improved modelling of dielectric properties, beyond the simplified tight-binding model used here, are likewise needed to better assess the properties of the films for such applications. In addition to a better electronic structure description, a more precise modelling of the dielectric function in the MHz to GHz range (including vibrational and intraband effects) would be of importance for comparison with experiments and to analyse the behaviour of $\rm \alpha$-BN films at device operation frequencies. Overall, the results confirm that amorphous boron nitride compounds are effective dielectric barriers against Cu diffusion, offering a promising strategy to reduce time delays in scaled interconnect architectures.
\section*{Acknowledgements}
This project has been supported by Samsung Advanced Institute of Technology and is conducted under the REDI Program, a project that has received funding from the European Union's Horizon 2020 research and innovation programme under the Marie Skłodowska-Curie grant agreement no. 101034328. This paper reflects only the author's view and the Research Executive Agency is not responsible for any use that may be made of the information it contains. This work was supported by Institute for Basic Science (IBS-R036-D1), Republic of Korea. ICN2 acknowledges the Grant PCI2021-122092-2A funded by MCIN/AEI/10.13039/501100011033 and by the “European Union NextGenerationEU/PRTR”. S. R. is also supported by MICIN with European funds NextGenerationEU (PRTRC17.I1) funded by Generalitat de Catalunya and by 2021 SGR 00997. H.-J.S. acknowledge support from the National Research Foundation of Korea (NRF) grant funded by the Korea government Ministry of Science and ICT (RS-2024-00352458). Simulations were performed at the Texas Advanced Computing Center (TACC) at The University of Texas at Austin and the Center for Nanoscale Materials, a U.S. Department of Energy Office of Science User Facility, supported by the U.S. DOE, Office of Basic Energy Sciences, under Contract No. DE-AC02-06CH11357.

\appendix

\section{Gaussian Approximation Potentials}
\label{a:GAP}
\subsection{Generation of the Training Data}
\begin{figure}[htb!]
    \centering
    \includegraphics[width=0.8\columnwidth]{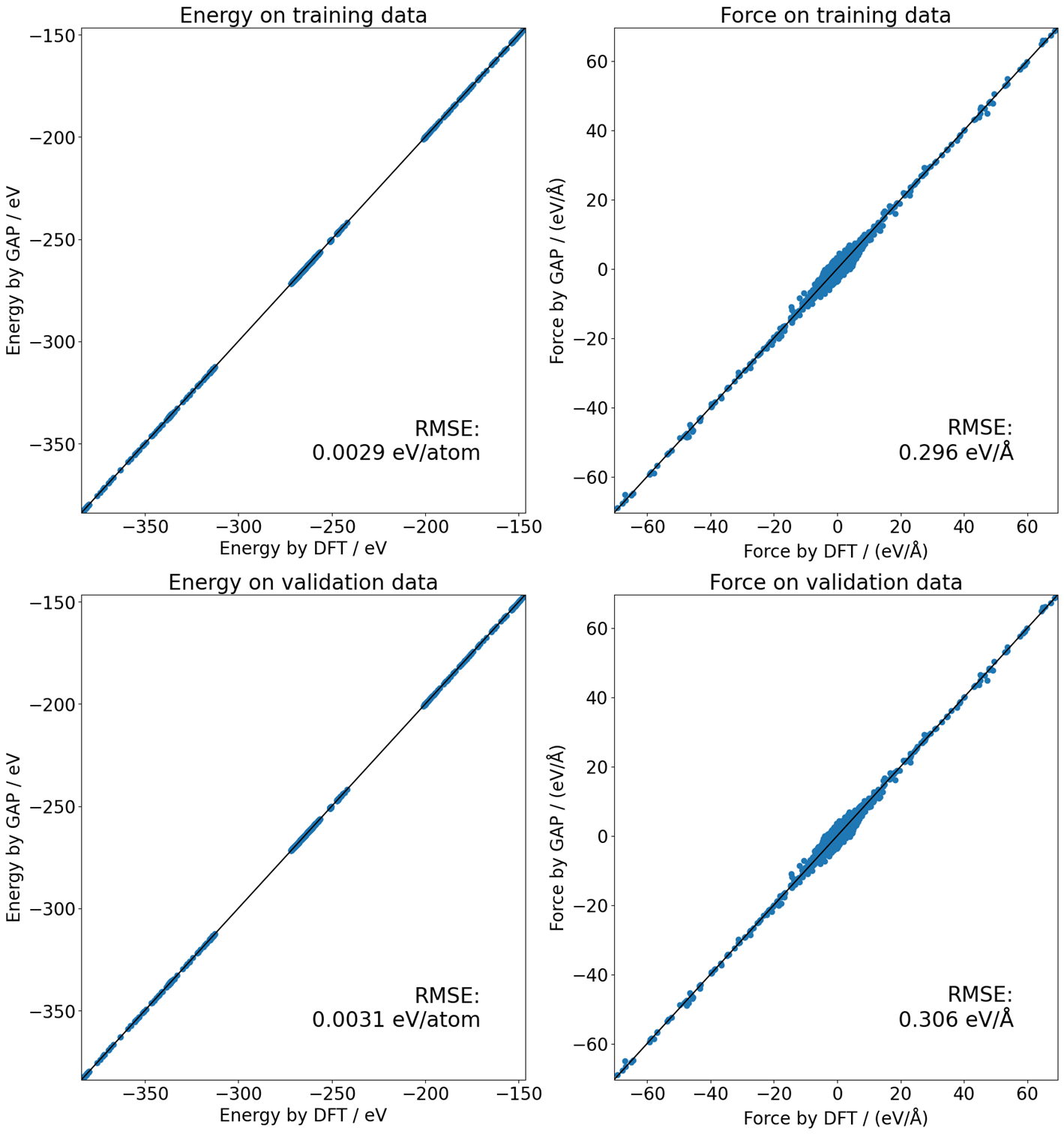}
    \caption{RMSE of energy and forces on training and validation sets. }
    \label{fig:rmse}
\end{figure}

The GAP model used in this study was developed using two datasets, referred to as the ``training'' and ``validation'' sets, which include atomic positions, energies, forces, and stresses calculated using Car-Parrinello molecular dynamics (CPMD) and density functional theory (DFT) with the Quantum ESPRESSO package~\cite{giannozzi2009quantum,giannozzi2017advanced,giannozzi2020quantum}. The training process followed an iterative scheme: (1) a large number of initial structures were generated by calculating the energies, forces, and stresses of molecular compounds, crystalline phases, and amorphous structures containing B, N, and Cu atoms; (2) an initial GAP model was trained on this dataset; (3) a new set of validation structures was generated using the previously trained GAP model; (4) the energy and force predictions of the GAP model were evaluated using DFT calculations. These training–validation cycles were repeated until the lowest RMSE values were achieved. Additionally, the accuracy of the final model was further assessed by comparing structural and energetic properties, as described below.\\
The ``training'' set used to generate and validate the final GAP model contains approximately 2500 structures, while the final validation set comprises 2000 structures. Together, the dataset includes around 200,000 atomic environments. A detailed breakdown by system type, number of atoms per configuration, and number of configurations is provided in Table~\ref{tab:dataset_summary}. The dataset spans a wide variety of configurations to ensure comprehensive coverage of relevant atomic environments. Initially, we incorporated several molecular compounds of B and N, as well as Cu systems, available in the Materials Project database~\cite{MaterialsProject}. Crystalline forms of BN and Cu were also included using the same source. These crystalline samples were expanded into larger supercells, and additional configurational diversity was introduced through the application of random defects and strain. Crystalline BN/Cu heterostructures were also generated to model interfacial environments. To cover disordered regimes, we generated a large number of liquid and quenched structures. These include $\alpha$-B, $\alpha$-Cu, $\alpha$-BN, $\alpha$-BN/Cu heterostructures, and randomly generated $\alpha$-B$_x$N$_y$Cu$_z$ configurations, where $x$, $y$, and $z$ are random values constrained by $x + y + z = 1$. All amorphous and liquid samples contain 64 atoms per unit cell, with cubic cell geometries. Cell parameters were randomly assigned to ensure that the resulting densities span the range 1.5–5.0~g/cm$^3$. Periodic boundary conditions were applied to all systems. For layered structures and heterostructures, a 10$\AA$ vacuum spacing was added along the $z$-direction; for isolated atoms and molecular compounds, vacuum was added in all three spatial directions.\\
\begin{table}[h]
\centering
\small
\begin{tabular}{lcc}
\toprule
\textbf{System} & \textbf{Num. of Atoms} & \textbf{Num. of Config.} \\
\midrule
Isolated elements (B, N, Cu) & 1 & 3 \\
Small molecules (B\textsubscript{2}, N\textsubscript{2}) & 2–16 & 100 \\
Crystalline phases (e.g.,h-BN, c-BN, fcc-Cu) & 4–64 & 1900 \\
Amorphous structures (e.g., $\alpha$-BN, $\alpha$-B\textsubscript{x}N\textsubscript{y}Cu\textsubscript{z}) & 64 & 2500 \\
\bottomrule
\end{tabular}
Breakdown of the training and validation datasets by system type, number of atoms per configuration, and total number of configurations used for GAP model development.
\label{tab:dataset_summary}
\end{table}
Car-Parrinello MD simulations for melt and quenched samples were performed using ultrasoft pseudopotentials (USPP) with the LDA exchange-correlation functional, employing kinetic energy cutoffs of 20 Ry for wavefunctions and 150~Ry for the charge density. DFT calculations for dataset labelling were performed using PAW pseudopotentials with the PBE exchange-correlation functional, with plane-wave cutoffs of 75 Ry for wavefunctions and 600 Ry for the charge density. To describe van der Waals interactions at BN/Cu interfaces, the non-local vdw-df-obk8 functional was employed~\cite{Klimes_2010}. When structures were relaxed, ionic steps were continued until all atomic forces were below 0.001 atomic units (a.u.). A Gaussian smearing of 0.1 eV was applied to the electronic levels during DFT calculations. $\Gamma$-point sampling was used for CPMD simulations, while Monkhorst-Pack grids (8$\times$8$\times$8 for three-dimensional systems and 8$\times$8$\times$1 for monolayers) were employed for DFT calculations.\\
The GAP model was trained using the QUIP software \cite{Csanyi2007-py, Bartok2010}. We employed three descriptors: two-body(2b), three-body(3b), and smooth overlap of atomic positions (SOAP). The 2b descriptor is used to describe the pairwise interactions between atoms. The 3b and SOAP descriptors represent the contribution of the atomic triplets and the local environment around an atom. The cut-offs and hyperparameters for each descriptor have been presented in Table \ref{table:gap_param}. The GAP models have been proven to be useful for several materials \cite{Kaya2023, kaya2023impact, Deringer2017, Deringer2021}. Sparsification using the CUR method \cite{mahoney2009cur} was selected for the SOAP kernel, while sparsification on a uniform grid was used for two-body and three-body kernels. The quality of the trained GAP model was later assessed by comparing the obtained energies per atom and forces of samples in the training and validation sets between the GAP model predictions and DFT calculations. The root-mean-square errors for energy and forces on training data are 0.0029 eV/atom and 0.296 eV/$\rm \AA$ and on validation data are 0.0031 eV/atom and 0.306 eV/$\rm \AA$, as shown in Fig. \ref{fig:rmse}. The obtained RMSE values are similar to the values in the literature for similar amorphous structures \cite{Sivaraman2020, Deringer2017, Fujikake2018}. Given that amorphous materials and complex systems, like the one we presented this study, require a highly diverse dataset, the RMSE values are considered as acceptable\cite{Fujikake2018}.\\
A slightly higher RMSE for the validation set is expected since those structures are not used for training the model. Low RMSE values show a good agreement between the GAP model and DFT calculations. \\
To further validate the potential, we calculated the average bond lengths of h-BN and Cu (111) slabs using both DFT calculations and GAP potentials by relaxing the structures. We also calculated the average interlayer distance between hBN and FCC Cu. For DFT relaxation, the same pseudopotentials and cutoffs used to generate the dataset were applied. The results show excellent agreement between DFT and GAP. The average bond length between B and N atoms in hBN was measured as 1.450 Å with GAP-MD and 1.447 Å with DFT, while the interatomic distance between Cu atoms was calculated as 2.540 Å using GAP-MD and 2.534 Å using DFT. These values are not only consistent with each other but also with reported values in the literature \cite{Shinde2020,Wolloch2014,Kaya2023}. The interlayer distance between hBN and Cu was calculated as approximately 3.327 Å using GAP-MD, while ab-initio MD simulations produced an interlayer distance of 3.332 Å. Reported interlayer distances between hBN and Cu in the literature range from 2.44 Å to 3.34 Å \cite{lyalin2014adsorption, Brulke2017, Feigelson2015, Lin2015,Koitz2013}. This variation can be attributed to differences in functionals, pseudopotentials, or software. However, both GAP-MD and ab-initio MD results fall within this range, demonstrating the reliability of the potential. \\

\begin{table}[htb!]
\begin{center}
    \begin{tabular}{cccc}
    \hline 
    & 2-body& 3-body& SOAP\\ \hline \hline
    $\delta$ (eV)& 2.0&0.1& 0.1\\ 
    $r_{cut}$ ($\rm \AA$)& 4.0& 3.2&7.0\\
    $r_{\Delta}$ ($\rm \AA$)& & & 0.5\\ \hline
    $\sigma_{at}$ ($\rm \AA$)& & &0.5\\
    $n_{max}$, $l_{max}$ & & &8, 8\\
    $\zeta$& & &4 \\ \hline
    Sparsification& Uniform & Uniform& CUR \\ \hline
    $N_t$ ($\rm \alpha$-BN/Cu samples)& & 150&3000\\
    $N_t$ (Crystalline samples)& & 50&1000\\
    $N_t$ (Total)& 25& 200&4000\\ \hline
    \end{tabular} 
    \caption{Parameters used to train the GAP potential for $\rm \alpha$-BN/Cu heterostructures.}
    \label{table:gap_param}
\end{center}
\end{table}

\clearpage

\section{Additional Morphological Information}

\label{a:pe}

\begin{figure}[htbp]
  \centering
  \begin{subfigure}[t]{0.48\textwidth}
    \centering
    \includegraphics[width=\linewidth]{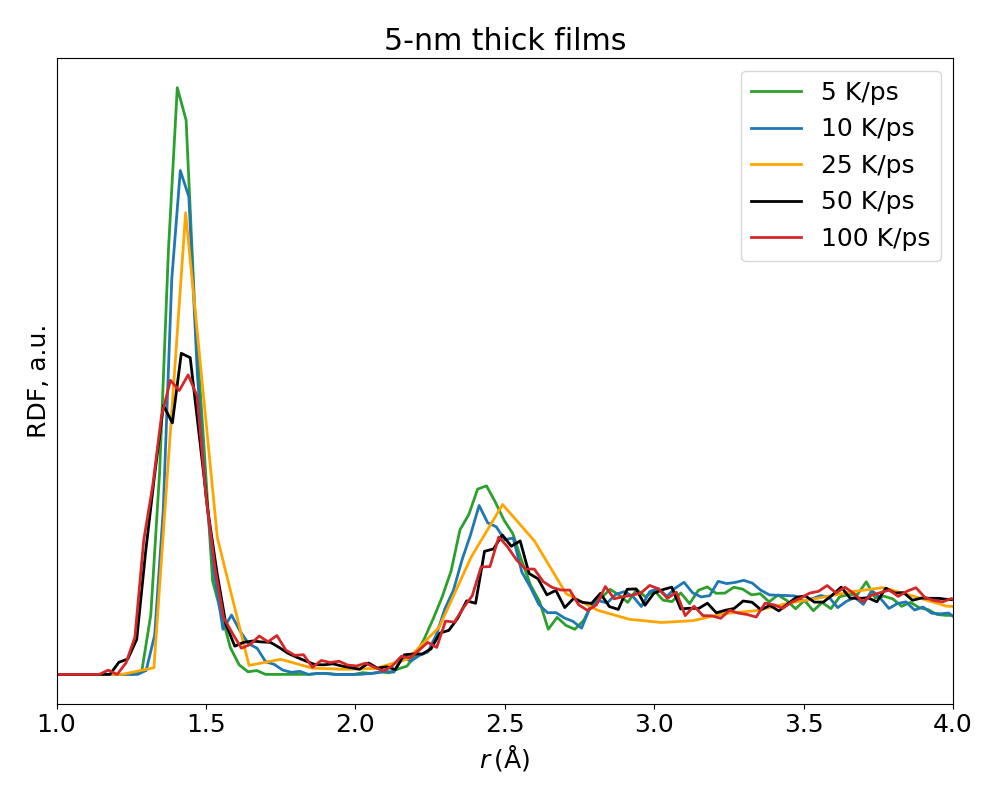}
    \caption{ }
    \label{fig:rdf_5nm}
  \end{subfigure}
  \hfill
  \begin{subfigure}[t]{0.48\textwidth}
    \centering
    \includegraphics[width=\linewidth]{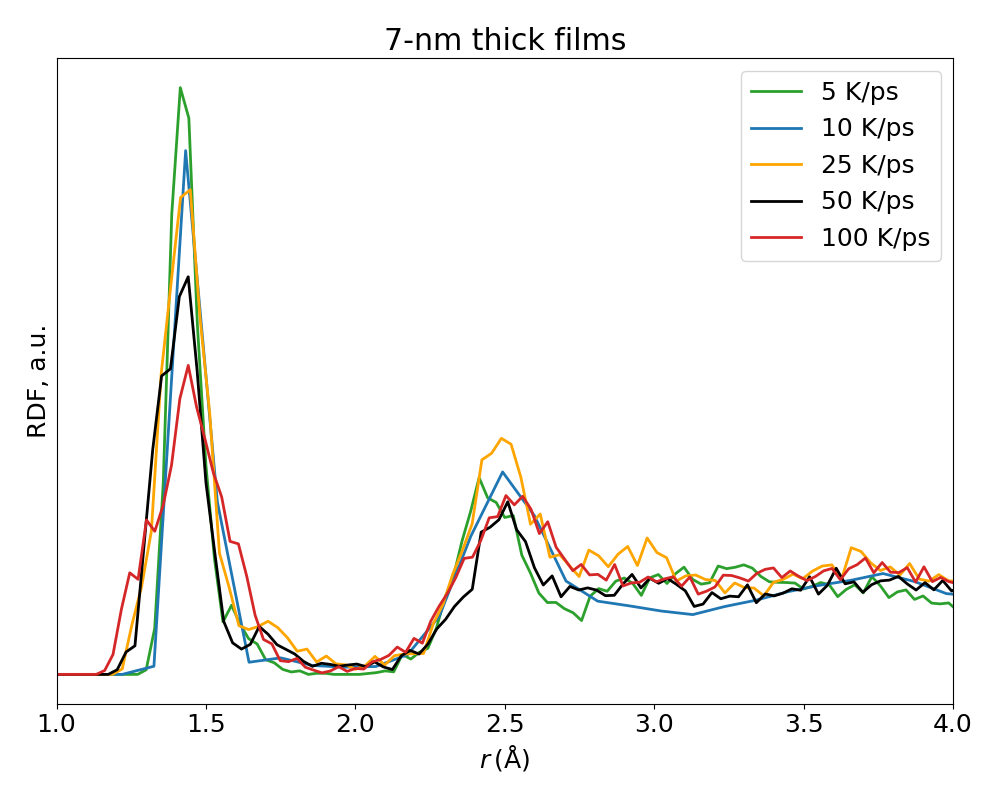}
    \caption{ }
    \label{fig:rdf_7nm}
  \end{subfigure}
  \caption{Radial distribution functions of 5 nm (a) and 7 nm thick (b) $\rm \alpha$-BN films at different cooling rates.}
  \label{fig:rdf_films}
\end{figure}

\begin{figure}[htbp]
  \centering

  \begin{subfigure}[t]{0.48\textwidth}
    \centering
    \includegraphics[width=\linewidth]{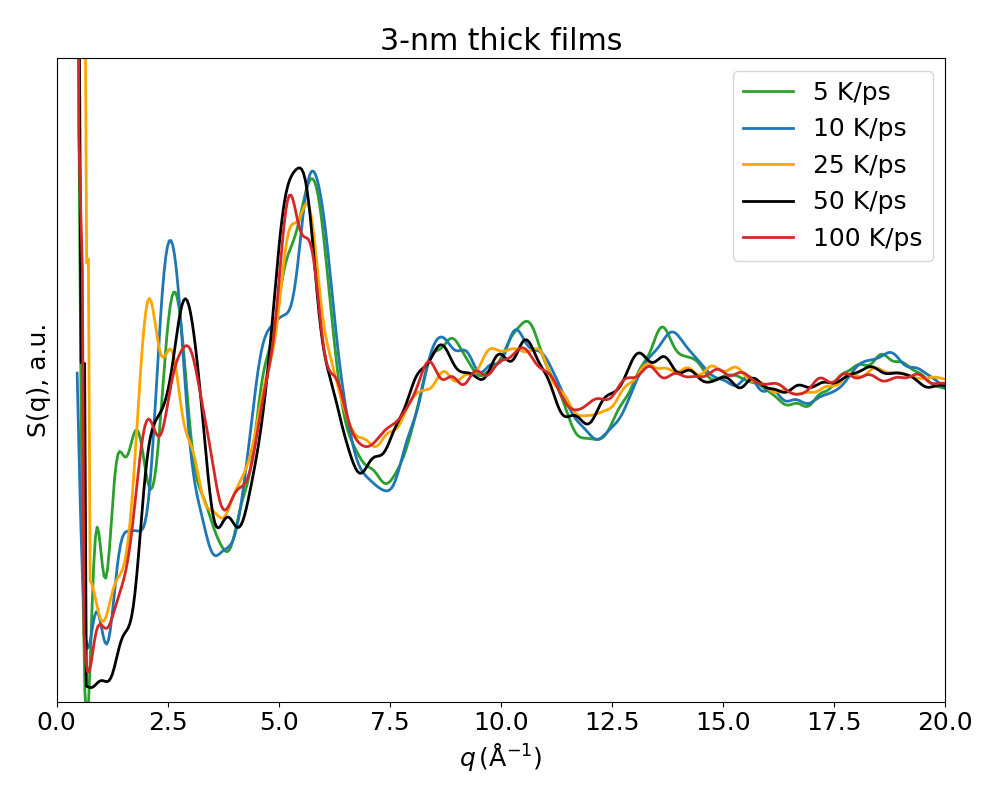}
    \caption{ }
    \label{fig:sq_3}
  \end{subfigure}\hfill
  \begin{subfigure}[t]{0.48\textwidth}
    \centering
    \includegraphics[width=\linewidth]{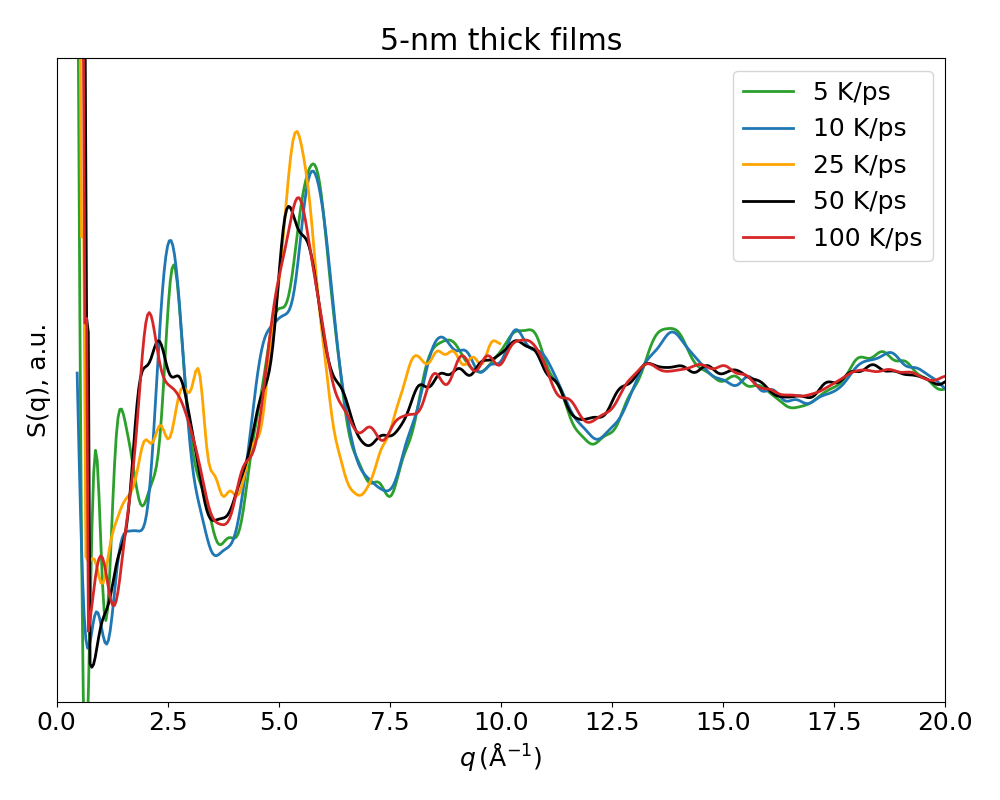}
    \caption{ }
    \label{fig:sq_5}
  \end{subfigure}
  \vspace{0.8em}
  \begin{subfigure}[t]{0.6\textwidth}
    \centering
    \includegraphics[width=\linewidth]{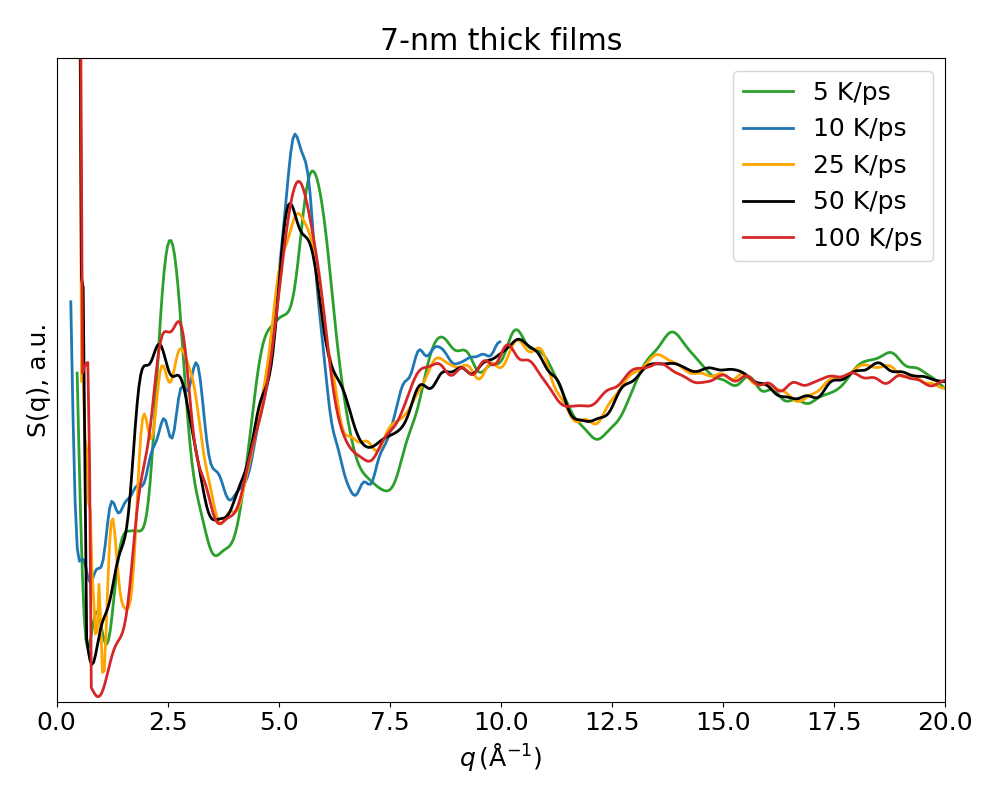}
    \caption{ }
    \label{fig:sq_7}
  \end{subfigure}
  \caption{Structure factors $S(q)$ of 3 nm (a), 5 nm (b), and 7 nm thick $\rm \alpha$-BN films generated at different cooling rates.}
  \label{fig:sq_triangle}
\end{figure}

\begin{figure}[h!tb]
    \centering
    \includegraphics[width=1.0\columnwidth]{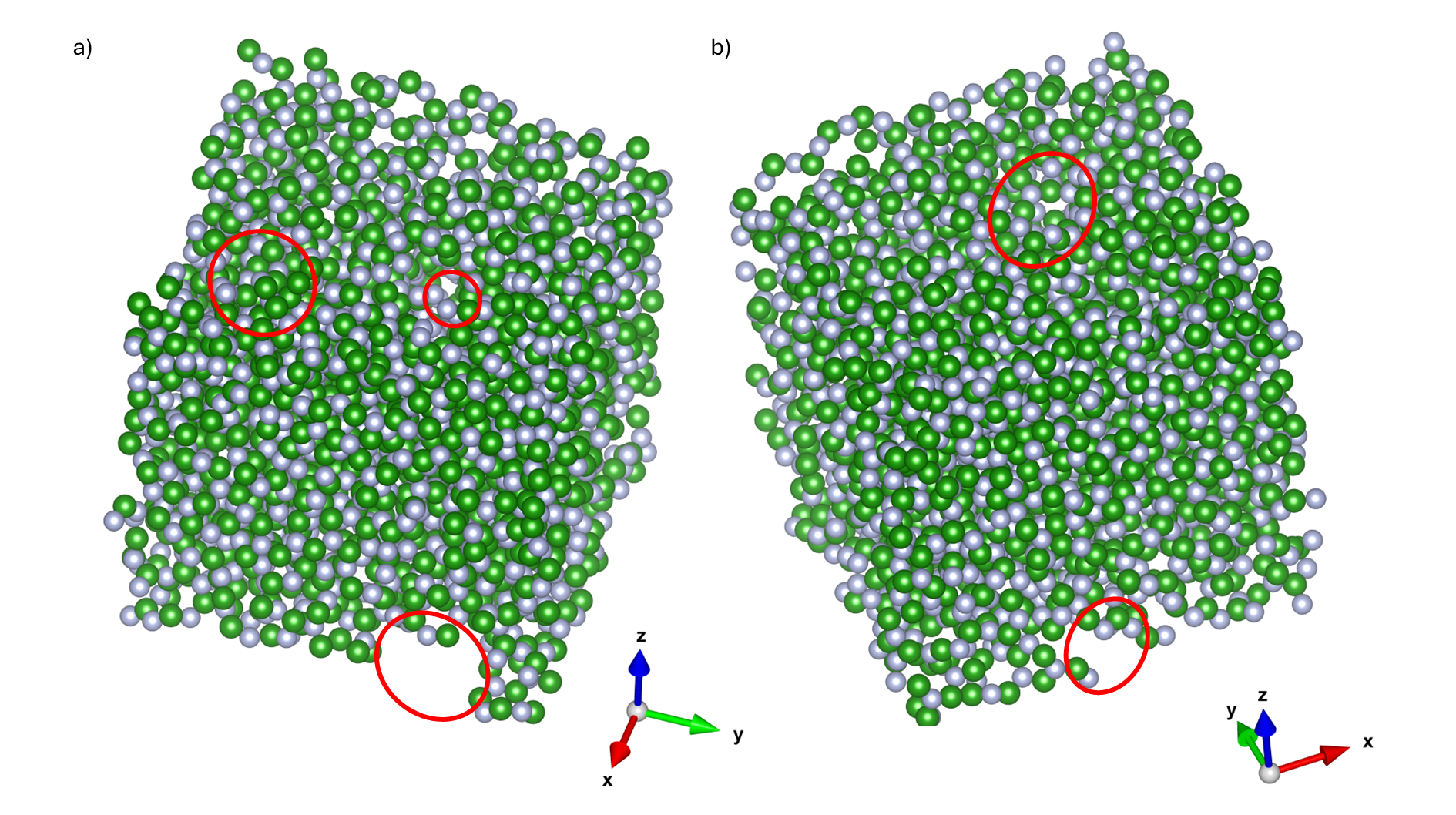}
    \caption{Nanovoids and pinholes observed in one of the 3-nm thick $\rm \alpha$-BN sample. Both images belongs to the same sample, they are tilted to show the nanovoids and pinholes within the structure. With high cooling rates, such defects were formed in $\rm \alpha$-BN samples. Large number of the atoms surrounding these nanovoids and pinholes are mostly $\rm sp^1$-hybridised (have coordination number:2).} 
    \label{fig:pores}
\end{figure}

\begin{figure}[htb!]
    \centering
    \includegraphics[width=0.8\columnwidth]{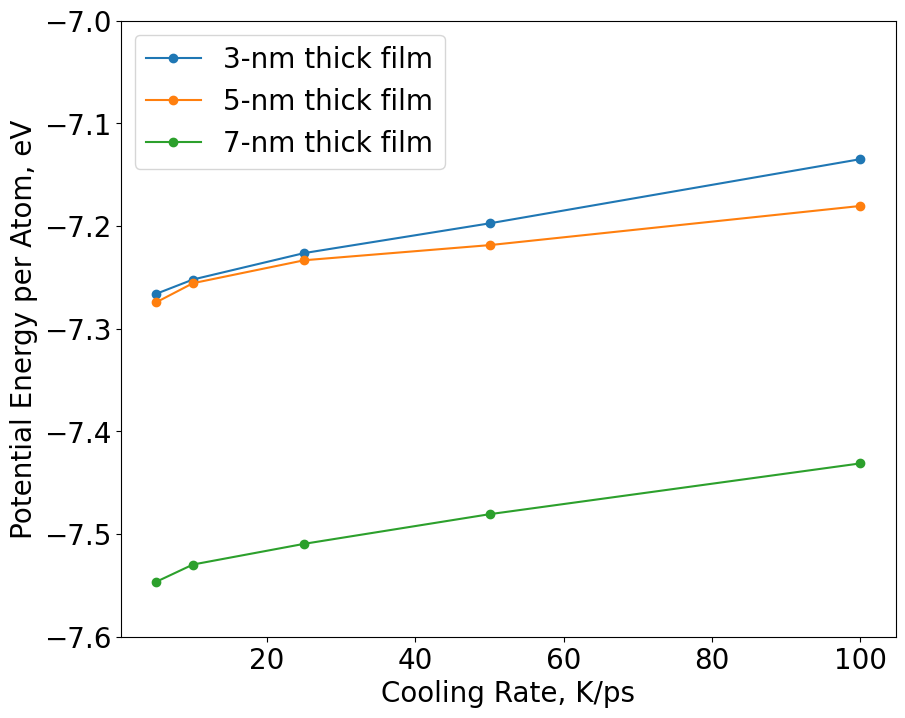}
    \caption{Potential energy per atom with respect to cooling rates and thicknesses of the films. This is a further parameter we used to evaluate the stability of the films. Moreover, the potential energy values increased with larger cooling rates that we applied to generate $\rm \alpha$-BN films. This visualises the impact of cooling rate on the stability of films.}
    \label{fig:pe}
\end{figure}

\clearpage

\section{Diffusivity vs Film Mass Density at High Temperatures}
\label{a:Dvsdensity}

\begin{figure}[htb!]
    \centering
    \includegraphics[width=\columnwidth]{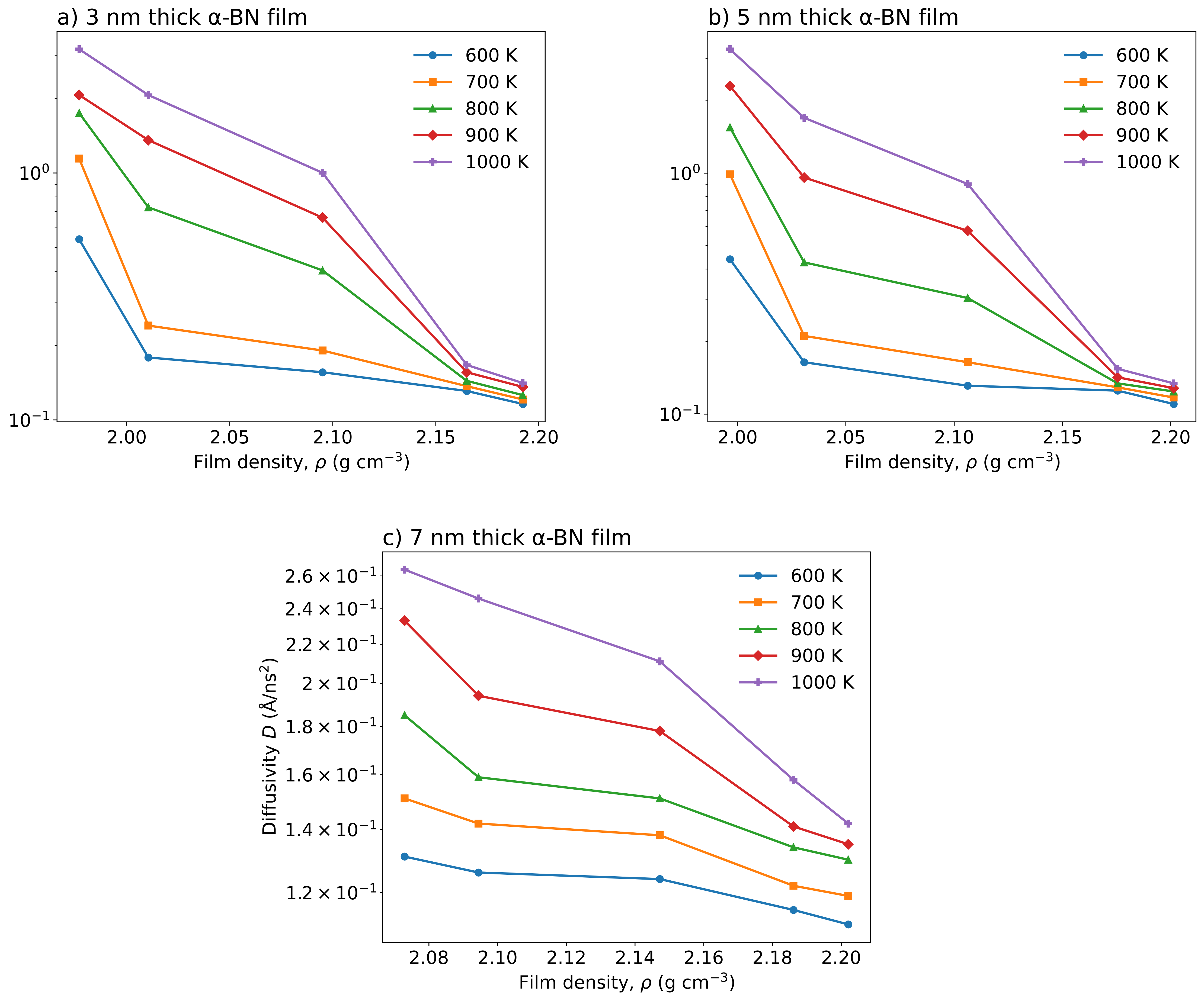}
    \caption{Diffusivity versus film mass density value at 300 K (after prepared). Panels correspond to 3, 5, and 7 nm films. The y-axis is $\log D$ ($\mathrm{\AA^2\,ns^{-1}}$) and the x-axis is $\rho$ ($\mathrm{g\,cm^{-3}}$). Density is computed from the slab volume $V = A\,t$ at the target temperature.}
\label{fig:Dvsrho}
\end{figure}

\clearpage

\section{SEM Images of $\rm \alpha$-BN Diffusion Barriers}
\label{a:sem}
\begin{figure}[htb!]
    \centering
    \includegraphics[width=1.0\columnwidth]{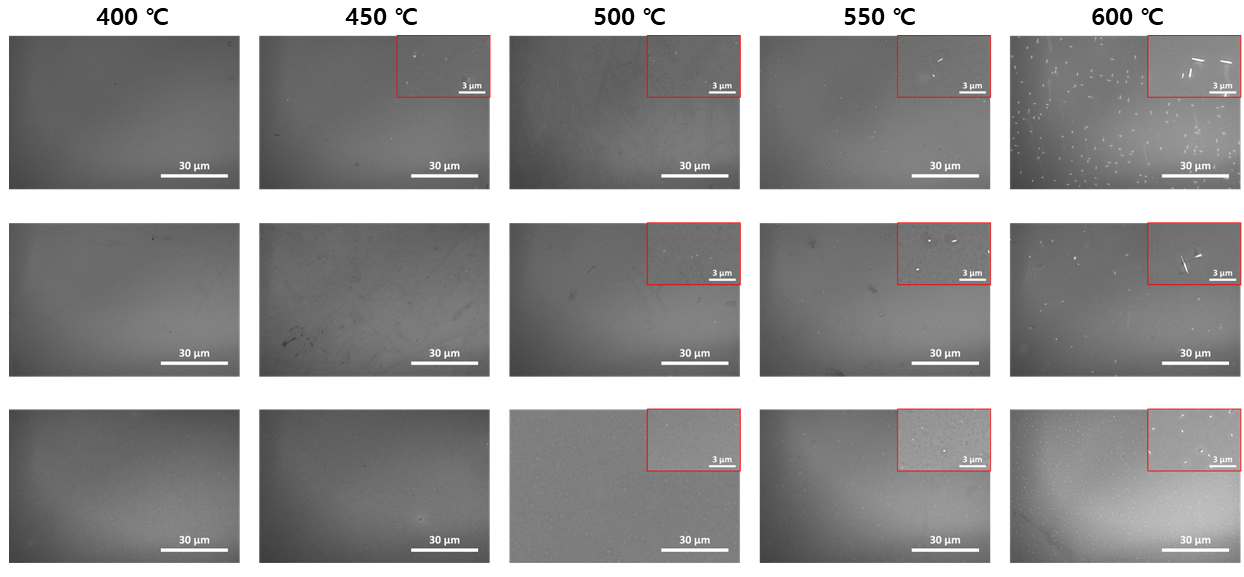}
    \caption{SEM images of 3-nm, 5-nm, and 7-nm thick $\rm \alpha$-BN/Si after 400 ~ 600 ℃ annealing and removing residual Cu films with Cu etchants. the insets boxed in red represent enlarged images of the copper silicide.}
    \label{fig:sem}
\end{figure}

\clearpage

\section*{References}
\bibliographystyle{iopart-num}
\bibliography{biblio}

\end{document}